\documentclass[pra,twocolumn,showpacs,floatfix]{revtex4-1}
\usepackage{times,amsmath,amssymb,amstext,latexsym,float,graphicx,color}
\usepackage{ulem,float,physics}
\usepackage[hidelinks]{hyperref}
\hypersetup{colorlinks=true, citecolor=blue, urlcolor=blue, linkcolor=blue}

\usepackage{soul}

\begin{document}
\setstcolor{red} 

\title{A two-state model for vortex nucleation in a rotating Bose-Einstein condensate}

\author{G.~Eriksson$^1$, J.~Bengtsson$^1$, G.~M.~Kavoulakis$^2$, and S.~M.~Reimann$^1$}
\affiliation{$^1$Mathematical Physics and NanoLund, LTH, Lund University, P. O. Box 118, SE-22100 Lund, Sweden}
\affiliation{$^2$Hellenic Mediterranean University, P. O. Box 1939, GR-71004, Heraklion, Greece}

\date{\today}

\begin{abstract} 
It is well-known that a rotating Bose-Einstein condensate forms vortices to carry the angular momentum. For a first vortex to nucleate at the trap center, the 
rotational frequency must become larger than a certain critical value. 
The vortex nucleation process, however, is sensitive to the trap shape.  
It was shown earlier~\cite{Dagnino2009a} that for a symmetry-breaking potential that preserves parity, at criticality the leading natural orbitals may become degenerate, giving rise to a ``maximally entangled" quantum state,  found from exact solutions for just a few bosons.  
Developing an effective two-state model, we here show that in the limit of large particle numbers,  the many-body ground state becomes either a so-called ``twin'' -like or a ``Schr\"odinger cat"-like state. 
We corroborate this finding by a direct comparison to the exact numerical solution of the problem, feasible for moderate particle numbers $N\lesssim 50$ within the lowest Landau level approximation. We show that the nature of the quantum state at criticality can be  controlled by both the quadrupolar  deformation and the flatness of the confining potential.
\end{abstract}
\maketitle

\section{Introduction}
When a harmonically trapped atomic Bose-Einstein condensate \cite{Pethick,Stringari} is set to rotate, quantized vortices may form to carry the angular momentum. These vortices are topological singularities characterized by  a phase jump around distinct density minima. The  spatial width of these minima is determined by the size of the healing length. With increasing rotation the number of vortices located within the condensate grows and eventually a vortex lattice forms~\cite{Chevy2000, Madison2000, Madison2001, Haljan2001, Hodby2001, Raman2001, Abo-Shaeer2001, Abo-Shaeer2002, Engels2002, Engels2003, Schweikhard2004}, as it is well-known for superfluids.
Another prominent  example are Helium nanodroplets~\cite{Barranco2006} that have also been visualized experimentally~\cite{Gomez2014}. 

The theoretical description of vortices and vortex lattices in Bose-Einstein condensates is often based on the Gross-Pitaevskii approach~\cite{Stringari1999,Butts1999, Linn1999, Feder1999, Kavoulakis2000, Linn2001, Garcia2001, Sinha2001, Kasamatsu2003, Vorov2005,Lieb2009}, but also methods going beyond mean-field were applied, see for example,~\cite{Wilkin1998, Mottelson1999, 
Bertsch1999, Kavoulakis2000, Jackson2000, Smith2000, Papenbrock2001, Huang2000, Jackson2001, Liu2001,Tsubota2002,Kasamatsu2003, 
Manninen2005, Reimann2006, Barberan2006, Dagnino2007, Cooper2008, Parke2008, Romanovsky2008, Liu2009, Dagnino2009a, 
Dagnino2009b, Papenbrock2012, Cremon2013, Cremon2015, Weiner2017,Beinke2018}. Especially 
in the  limit of rapid rotation,  where pure mean-field fails and the bosonic cloud resembles a (properly symmetrized) Laughlin state~(see for example ~\cite{Wilkin1998,Cooper1999,Wilkin2000,Viefers2000} and the reviews \cite{Viefers2008, Cooper2008}). Extensive reviews on the physics of rotating condensates are also found in Refs.~\cite{Bloch2008, Fetter2009, Saarikoski2010}. 

For a dilute and weakly interacting harmonically trapped Bose gas, 
the nucleation process of the first vortex (the so-called `unit vortex''~\cite{Mottelson1999,Bertsch1999,Papenbrock2001}) with increasing angular 
momentum $L= M_L\hbar$ 
was associated with a Nambu-Goldstone mode that becomes massive when $M_L$ equals to the number of bosons $N$ in the system~\cite{Ueda2006}.  During the nucleation process, the system undergoes a transition from a condensate in the single-particle  orbital with zero angular momentum to a state that has a macroscopic occupancy in the single-particle orbital with angular momentum $\hbar$.
The detailed description relating the emergence of this first vortex when passing a critical  
rotation frequency $\Omega _c$, however, is intimately connected with the symmetry of the trap~\cite{Parke2008, Dagnino2009a, Dagnino2009b}. For systems with an even number of bosons it was found that a small quadrupole deformation (opening a small energy gap between the ground- and first excited state of the system as opposed to the azimuthally symmetric case) gave rise to a density matrix with {\it two} dominant natural orbitals at criticality. These two orbitals are of different parity and  have  comparable macroscopic occupancies~\cite{Dagnino2009a, Dagnino2009b} which together almost make it up to $N$; in other words, the many-body ground state at $\Omega_c$ is fragmented~\cite{Penrose1951, Penrose1956, Yang1962}. 
Despite the well-known exactness of the Gross-Pitaevskii expansion of the energy in terms of $1/N$ in the thermodynamic limit~\cite{Lieb2009} (valid for a dilute, weakly interacting single-component gas at moderate rotation), at a rotation frequency of  $\Omega =\Omega _c$ 
 even a perturbatively small trap deformation may thus render the description of the ground state by  a single order parameter insufficient. 
On either side of the critical frequency, however, the respective single order parameter obtained from the mean-field approach rather accurately describes the ground state, with its structural change reflecting the change of symmetry. 
For even $N$ and sufficiently weak interactions, it was  suggested that at quantum criticality, {\it i.e.} at $\Omega =\Omega _c$, the two modes give rise to 
a maximally entangled state, proposed as a superposition $(\lvert N,0 \rangle + \lvert N-2, 2 \rangle +\dots +\lvert 0, N \rangle ) /\sqrt{N/2+1}$ resulting from the parity-conserving quadrupole deformation of the trap~\cite{Dagnino2009a, Dagnino2009b}. Here $\lvert n_1,n_2\rangle$ is the correctly symmetrized many-body state obtained with $n_1$ and $n_2$ bosons in the two considered natural orbitals of different parity, respectively. In the limit of small (and even) $N \lesssim 20$ this maximally entangled state was attributed a larger overlap with the exact many-body ground state compared to ``Schr\"odinger cat''- or ``twin''-like states.  
While ``cat''-like states are of the form $(\lvert N,0 \rangle + \lvert 0,N \rangle )/\sqrt{2}$, being a superposition of two states fully condensed in either of the two modes, ``twin'' states have equal occupation in each mode, $\lvert N/2, N/2\rangle $. 
We note here that such cat states have been proposed before, but for small $N$ and stronger interactions~\cite{rico2013}. 

Given the recent interest in correlated macroscopic quantum states~\cite{Frowis2018}, it is interesting to revisit the fate of the maximally entangled state~\cite{Dagnino2009a, Dagnino2009b} for weakly interacting gases at criticality in the limit of larger particle numbers, which is the main theme of this article. We develop an effective two-state model for the many-body problem right at criticality. A semiclassical analytic solution allows us to extract the nature of the transition to a vortex-carrying state for a large atom number. We also compare the results of our model with the full quantum mechanical solution. The latter is naturally restricted to relatively small particle numbers due to the increasing complexity of the problem. Nevertheless, with increasing $N$ we see a clear trend favoring the ``twin" or ``cat"-like states rather than the ``maximally entangled" state of Refs.~\cite{Dagnino2009a,Dagnino2009b}. 

\section{Correlated states in the process of vortex nucleation}
\label{sec:CI}

Let us  recapitulate how the nucleation of the unit vortex 
is uncovered in the structure of the exact eigenspectra or in the natural orbitals. 
We consider an even number $N$ of spinless bosons in a quasi two-dimensional 
harmonic trap with $\omega = \omega_x = \omega_y \ll \omega _z$ that revolves about the $z$-axis with a constant angular frequency $\Omega $. The elastic atom-atom collisions are taken to be of  s-wave type,  modelled by the interaction potential $V_{\mathrm{int}}=g\delta ({\bf r}_i-{\bf r}_j)$ with interaction strength $g=4\pi \hbar ^2aM^{-1}\int \lvert \phi _0(z)\rvert ^4 dz$. (Here,  
$a$ is the 3D scattering length, $M$ the atom mass and $\phi _0(z)$
the single-particle oscillator ground state  in the tightly confined $z$-direction).  
In what follows below,  we set $\hbar = M=\omega =1$. 
When the system is dilute and weakly interacting (such that the typical interaction energy is much smaller than the oscillator quantum of energy), the effectively two-dimensional rotating condensate can be well described within the lowest Landau level~\cite{Mottelson1999,Ho2001,Morris2006} where a practically exact numerical solution can be obtained by brute-force diagonalization. The implied basis size restriction  conveniently leads to an effective short-range cut-off and thus implicitly regularizes the contact interaction~\cite{Rontani2017}. Care must however be taken that deviations from the trap harmonicity as well as the parameter $gN$ are sufficiently small to ensure that the lowest Landau level can still capture the complexity of the many-body state in question~\cite{Morris2006,rico2013}.
This approach has been extensively used in the past, see the reviews~\cite{Viefers2008,Cooper2008,Fetter2009,Saarikoski2010} 
or, {\it e.g.}~\cite{Cremon2013, Cremon2015} (and refs. therein).

For values of the total angular momentum $L$ 
in the range  $2 \le L \le N$ and for $L=0$  
the exact ground state energy in the rotating frame 
$E_{\mathrm{rot}} = N+L + gN (2N-L-2)/(8\pi)- \Omega L$  
is linear in $L$ even in the presence of interactions~\cite{Bertsch1999,Jackson2000,Papenbrock2001}. Consequently, there is a critical rotational frequency $\Omega _c=1-gN/(8\pi )$ of the trap which makes the energy of all yrast states with  $2 \le L \le N$ and the one with $L=0$ degenerate in the rotating frame.
(The exception, $L=1$, is a center of mass excitation from $L=0$).  
For $gN \ll 4\pi$ the lowest Landau level is  expected to be adequate. 

\begin{figure}[H]
\includegraphics[width=0.98\linewidth]{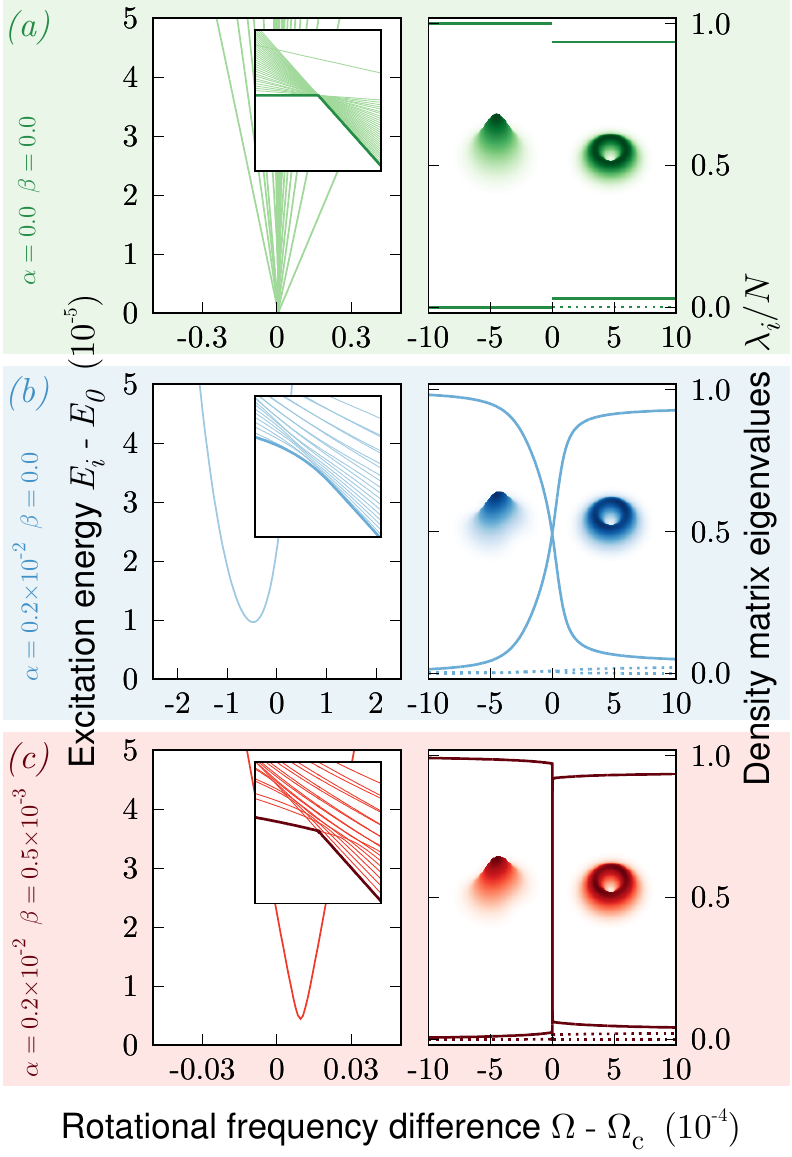}
\caption{{\it (Color online)} Energy gap of excitations relative to the ground state, $E_i-E_0$, in the rotating frame {\it(left panels)} and the corresponding eigenvalues of the  ground state single-particle density matrix {\it (right panels)} as a function of  $\Omega -\Omega _c$, {\it i.e.}, the difference with respect to the critical frequency  $\Omega _c$, for $N=30$ and $gN=1.5$.
Three different forms of the confinement are considered in {\it (a)-(c)}, as specified by the values of $\alpha $ and $\beta $ given in each panel. Note the different scales on the {\it x}-axis in the panels to the left.  (For reference, the corresponding many-body spectra are also shown as insets, 
 centered around $\Omega_c$ and $E_0(\omega _c)$ for the intervals 
 $E_i-E_0 (\Omega _c) \le 0.015$ and $\lvert \Omega -\Omega _c\rvert \le  0.0005$, which are omitted in the figure for simplicity. The ground state being marked by a thicker solid line). Clearly, the very weak quadrupolar and anharmonic contributions lead to a very small energy gap  (avoided crossing).  In the plots to the right, the two largest density matrix eigenvalues are plotted as a solid line, and all other ones by dashed lines. 
The density distributions are shown as insets on the left and right hand side of the transition to the first singly quantized vortex.  
}
\label{Fig1:Spectra}
\end{figure}

Figure~\ref{Fig1:Spectra} shows the (numerically exact) low-lying  excitation energies $E_i - E_0$ {\it (left)} and the density matrix eigenvalues {\it (right)} 
 of a harmonic trap with $N=30$ bosons as a function of the trap rotation $\Omega $ when $gN = 1.5$. ($E_0$ is the ground state energy at given $\Omega $) . {\it Panel (a)} is for an azimuthally symmetric trap, while for {\it  (b)} a quadrupole deformation was considered,  to which a  trap anharmonicity was added in {\it (c)}.   
The insets in the left panels show the structure of the many-body energies $E_i$, and the ones in the right panel show the density distributions on either side of criticality. 
We first consider the azimuthally symmetric case, shown 
in Fig.~\ref{Fig1:Spectra}(a).
At a certain critical rotation frequency $\Omega _c$ (which due to the weak interactions chosen here occurs at a value rather close to the trap frequency), as a consequence of the ground state degeneracy 
the system makes a discontinuous transition in  the angular momentum from $\ell =L/N=0$ for $\Omega \rightarrow \Omega _c^-$ to $\ell = 1$ for $\Omega \rightarrow \Omega _c^+$, where a single-quantized vortex is localized at the trap center. For a symmetric trap this transition is marked by a crossing of  
the many-body energy levels at $\Omega _c$ (see Fig.~\ref{Fig1:Spectra}(a), {\it  left}) associated with a discontinuous transition in the largest occupancies of the natural orbitals (see Fig.~\ref{Fig1:Spectra}(a), {\it right}). The insets show how the single-particle density of the ground state transforms from a Gaussian profile at slow rotation to a vortex, localized at the trap center,  beyond criticality. 

A small perturbation $ \alpha ( x^2- y^2)$ adds a weak  parity-conserving quadrupole deformation to the harmonic trap,  
as discussed in~\cite{Parke2008,Dagnino2009a,Dagnino2009b,rico2013}. The parameter $\alpha >0$ is here chosen sufficiently small such that the approximation to restrict the space to the lowest Landau level is not violated.  The critical frequency $\Omega_c$ now takes on a slightly different numerical value compared to that of the azimuthally symmetric harmonic trap. 
For such  a quadrupolar perturbation and  
for even $N$, the many-body spectrum  now exhibits an avoided level crossing at $\Omega _c$.   The corresponding excitation energies and density matrix eigenvalues are shown Fig.~\ref{Fig1:Spectra}(b). Note that the degeneracy 
at criticality is  lifted.
Albeit the gap at the avoided crossing is tiny  (owing to the smallness of $\alpha $),  the increase in the expectation value of $L$ occurs less abruptly compared to the azimuthally symmetric case, 
smoothening the transition at (and around) criticality. 
The many-body Hamiltonian conserves parity even with the quadrupole deformation switched on.  For even $N$ and below criticality, the dominant natural orbital has even parity, whereas it has odd parity above the transition. As seen in Fig.~\ref{Fig1:Spectra}(b) ({\it right }), 
at criticality the occupancies of the two most significant natural orbitals (with opposite parity) become of equal magnitude, implying that the state is fragmented~\cite{Parke2008,Dagnino2009a,Dagnino2009b}.  

When adding a further parity-conserving perturbation of the form  $\beta (x^2+y^2)^2$ 
that renders the potential slightly anharmonic for small $\beta > 0$ (again chosen small enough to stay within the lowest Landau level) we find that this transition between the leading natural orbitals becomes very narrow, as shown in Fig.~\ref{Fig1:Spectra} (c). 
For all parity-conserving deformations and for even $N$, the crux of the matter  lies in the fact that 
with {\it two} degenerate and macroscopically occupied natural orbitals instead of the usual single one,  the description with a single order parameter  fails to correctly describe the transition. In other words,  the usual Gross-Pitaevskii approach that correctly describes the non-rotating ground state as well as the unit vortex, cannot account for the correlations built up at criticality. 
In this context it is also instructive to briefly cast an eye on the structure of the Gross-Pitaevskii order parameter on the left and right hand side of the transition. 

In the appendix, we evaluate analytic mean-field results for the weak quadrupole symmetry-breaking potential,   identifying the relevant leading contributions of the single-particle states on either side of criticality. 
For  $\ell \rightarrow 0^+$ the order parameter, as in~\cite{Kavoulakis2000},  is a linear superposition $\psi ^{(0^+)}\approx c_0\phi_0 +c_2\phi_2$ with $\lvert c_0\lvert ^2 = 1- \ell/2$ and $\lvert c_2\lvert ^2 =\ell /2$ and $\phi _m=r 
^m {\mathrm e}^{(i m \vartheta - r^2/2)}/\sqrt{\pi m!}$ (in polar coordinates $r$ and $\vartheta $, where $m$ is the single-particle angular momentum quantum number).  
In the limit $\ell \rightarrow 1^-$ the order parameter  $\psi ^{(1^-)}\approx \phi _1$ plus corrections of order $\alpha ^2 /(gN)^2$ (where this correction is referring to the occupancy of the $m=3$ state).
While the Gross-Pitaevskii solution correctly describes the symmetry of the full solution offside criticality, we have seen above that a single order parameter 
cannot fully capture the state across the transition, where the exact solution is represented mainly by two equally populated natural orbitals of different parity. 
In the following we thus develop  an effective two-state model that 
in the limit of large (yet, even) $N$ would allow to assess the structure of the state right at the critical frequency $\Omega _c$.

\section{Two-state model at quantum criticality}
\label{sec3}

As remarked in Refs.~\cite{Dagnino2009a,Dagnino2009b} and also discussed above, see Fig.~\ref{Fig1:Spectra}, for a small quadrupolar symmetry breaking perturbation one finds that at criticality, 
the two largest density matrix eigenvalues  are $\lambda _1= \lambda _2= N/2$. This indicates that it may be sufficient to describe the system  in terms of the correctly symmetrized many-body states $\lvert \psi^{N-n}_1\psi^n_2\rangle$, or equivalently $\lvert N-n,n \rangle$ in the occupation number representation of the natural orbitals (density matrix eigenstates) $\lvert \psi _1\rangle $ and $\lvert \psi _2\rangle $  of different parity, corresponding to  $\lambda _1$ and $\lambda _2$, respectively. (We emphasize again that due to parity conservation by the perturbation for even $N$ only states with even $n$ contribute to the many-body ground state). 

Motivated by the equality of the two largest density matrix eigenvalues at criticality in the exact solutions,  and likewise by the simple structure of the Gross-Pitaevskii  order parameter offside the transition, let us now develop a two-state model at $\Omega _c$ in order to try to approximately capture the large-$N$ limit of the entangled state at criticality. We hereby make use of a version of the Lipkin-Meshkov-Glick (LMG) model~\cite{Lipkin1965} that was introduced already back in 1965 to describe phase transitions in nuclei but has then found applications in many different fields of physics (such as, for example, atomic Bose gases~\cite{Elgaroy1999} and the description of Josephson junctions~\cite{JuliaDiaz2013}).
After first establishing the solutions of this model analytically, we compare its predictions for the limit of large $N$ against results obtained by diagonalizing the many-body Hamiltonian matrix within the lowest Landau level discussed above,  which remains a viable approach in the limit of very weak interactions and not too large $N$.

At criticality, we  use the two dominant natural orbitals 
$\lvert \psi _1\rangle $ and $\lvert \psi _2\rangle $ to construct a reduced Hilbert space $\widetilde {\cal H}$ in which a substantial part of the many-body ground state $\lvert \Psi (\Omega _c)\rangle $ resides. For a non-rotating system, the Hamiltonian of the two-state model in the spirit of ~\cite{Lipkin1965} and ~\cite{Elgaroy1999} is  written as 
\begin{eqnarray}
\hat {\widetilde {H}}_{\mathrm{LMG}}&=&\varepsilon _1~a_1^{\dagger } a_1 ~  + 
\varepsilon _2~a_2^{\dagger } a_2
\nonumber \\
 &+& {1\over 2} ~V_{1111}~a_1^{\dagger }a_1^{\dagger }a_1a_1   +
        {1\over 2} ~V_{2222}~a_2^{\dagger }a_2^{\dagger }a_2a_2 \nonumber \\
 &+& ~2~V_{1212}~a_1^{\dagger }a_1a_2^{\dagger }a_2\nonumber \\
 &+& ~V_{1122}~a_1^{\dagger }a_1^{\dagger }a_2a_2  
        + ~V_{2211}~a_2^{\dagger }a_2^{\dagger }a_1a_1~.
        \label{2stateHamiltonian}
\end{eqnarray}
Here, as usual, the operators $a_j^{\dagger }$ and $a_j$ create and annihilate  quanta in the states $\lvert\psi_j\rangle$ (where $j=1,2$), $\varepsilon _j$  is the single-particle energy associated with the natural orbital $\lvert \psi _j\rangle$,
\begin{eqnarray}
\varepsilon_j &=& \int {\mathrm d}x {\mathrm dy}~\psi^*_j \Big[ -\nabla^2/2 + (x^2+y^2)/2 + \alpha (x^2 - y^2) \nonumber\\ &+&  \beta (x^2+y^2)^2  \Big] \psi_j~,
\end{eqnarray}
    and  $V_{ijkl}=g\int {\mathrm d}x{\mathrm d}y~\psi _i^*\psi _j^* \psi _k \psi _l$, where $i,j,k,l\in \{1,2\}$.

If setting the last two terms in Eq.~(\ref{2stateHamiltonian}) to zero one retrieves the Hamiltonian suggested by Nozi\`eres and Saint James~\cite{Nozieres1982} for particle condensation in a structureless Bose liquid.
In their model the exchange interaction implies that the state cannot be fragmented, provided that the sign of the coupling constant is positive. 
The last two terms in Eq.~(\ref{2stateHamiltonian}), absent in the approach of Ref.~\cite{Nozieres1982}, here make the crucial difference, since they enable the transfer of atoms from one orbital to the other, and a richer physical picture emerges. (Having in mind the structure of the Gross-Pitaevskii  order parameter on either side of the transition, it is instructive to see that they origin from processes where 
two atoms with, for example, $m=0$ and $m=2$ end up in an orbital with $m=1$, or vice-versa).

While the Hamiltonian, Eq.~(\ref{2stateHamiltonian}), in principle may straightforwardly be diagonalized numerically, we here choose to follow an analytical approach similar to Ref.~\cite{Elgaroy1999} that also holds in the limit of large $N$.

The two-state model can be described by the SU(2) algebra of ordinary spin $1/2$.  With the pseudo-spin operators 
\begin{equation}
 {\hat J}_+ = a_2^{\dagger} a_1, \,\, {\hat J}_- = a_1^{\dagger} a_2, \,\, {\hat J}_z = \frac 1 2 (a_2^{\dagger} a_2 - a_1^{\dagger} a_1)
\end{equation}
and the particle number operator $\hat N= a_1^{\dagger }a_1 + a_2^{\dagger } a_2$, one writes
\begin{equation}
a_1^{\dagger }a_1={{\hat N}\over 2}  -  {\hat J}_z~,\qquad a_2^{\dagger }a_2={{\hat N}\over 2}  +  {\hat J}_z~.
\end{equation}
Similarly to ~\cite{Elgaroy1999} we rewrite the Hamiltonian Eq.~(\ref{2stateHamiltonian}) in the pseudo-spin operators %
\begin{eqnarray}
\hat{\widetilde{H}}_{\mathrm{LMG}} &= &{{\hat N}\over 2} (\varepsilon _1+\varepsilon _2) + 
{\hat J}_z (\varepsilon _2-\varepsilon _1) +\nonumber \\
&+&{1\over 2} V_{1111}\biggl[{\hat N\over 2} \bigl( {\hat N\over 2}-1 \bigr) - 
\bigl(\hat N -1\bigr) \hat J_z + \hat J_z^2\biggr] \nonumber \\
&+&{1\over 2} V_{2222}\biggl[{\hat N\over 2} \bigl( {\hat N\over 2}-1\bigr) + 
\bigl(\hat N -1\bigr) \hat J_z + \hat J_z^2\biggr] \nonumber \\
&+& V_{1212} \bigl(\hat J_+\hat J_- + \hat J _-  \hat J_+ -\hat N\bigr) 
\nonumber \\
&+& V_{1122} \bigl(\hat J_+\hat J_+ + \hat J _-  \hat J_- -\hat N\bigr)~.
\label{hpseudospin}
\end{eqnarray}
Of course, with $\hat J _{\pm } = \hat  J_x \pm i \hat J _y$, $\hat {\widetilde {H}}_{\textrm{LMG}}$  can also be expressed in the pseudospin components 
$\hat J_x, \hat J_y$ and $\hat J_z$. Note also that $[\hat {\widetilde {H}}_{\mathrm{LMG}}, \hat J^2]=0$.
The states of the LGM model correspond to points on the Bloch sphere~\cite{Elgaroy1999,JuliaDiaz2013}. In the semiclassical approximation of large $N$, 
\begin{eqnarray}
\hat J_x &\rightarrow  & \frac N 2 \sin \theta \cos \phi \nonumber \\
\hat J_y &\rightarrow  & \frac N 2 \sin \theta \sin \phi \label{quasispin} \\
\hat J_z &\rightarrow  & \frac N 2 \cos \theta ~,\nonumber
\end{eqnarray}
where $\theta $ and $\phi $ are the corresponding spherical coordinates  of the Bloch sphere with radius normalized to $N/2$~\cite{Elgaroy1999}.

Now, for a rotating system in the rotating frame,
\begin{equation}
\hat {\widetilde {H}} _{\mathrm{rot}} = \hat {\widetilde {H}}_{\mathrm{LMG}} - \widetilde \Omega \sum _{j=1}^2 l _j~a_j^{\dagger } a_j ~, 
\label{rotframe}
\end{equation}
where $l _j=  \int {\mathrm d}x{\mathrm d}y~\psi _j^*  \hat L_z\psi _j$ and where $\hat{L}_z$ is the ($z$-component of the) angular-momentum operator. For the sake of generality, we here allow for a rotational frequency $\widetilde{\Omega}$ that differs from $\Omega_c$ (despite the fact that $\lvert \psi_1\rangle$ and $\lvert \psi_2\rangle$ are obtained for $\Omega_c$). Rewriting Eq.~(\ref{rotframe}) in terms of the above quasi-spin operators Eqs.~(\ref{quasispin}) in their semiclassical approximation, and as in~\cite{Elgaroy1999} taking $N/2(N/2-1)\approx N^2/4$ and $N-1\approx N$, 
one finally arrives at an expression for the semiclassical energy:
\begin{eqnarray}
\widetilde{E}_{\textrm{rot}} &\rightarrow & \frac{N}{2}\left(\epsilon_1 + \epsilon_2\right) + \left( \frac{N^2}{8} -\frac{N}{4}  \right) \left(V_{1111} + V_{2222}\right) 
 \nonumber \\ 
 & - & N V_{1212}  + \frac{N^2}{4} \left(2 V_{1212} + V_{1122}\cos 2\phi \right) 
\nonumber \\   
   & +&\frac{f_1(\phi)}{2}  \cos \theta  +  \frac{f_2(\widetilde{\Omega})}{2} \cos^2 \theta~,
\label{lipf}
\end{eqnarray}
where
\begin{equation}
f_1(\phi) =  \frac{N^2}{4}\left[V_{1111} + V_{2222} - 4 V_{1212} - 2 V_{1122}\cos2\phi\right],
\label{cala} 
\end{equation}
and 
\begin{eqnarray}
f_2(\widetilde{\Omega})  &=& \frac{N(N-1)}{2}\left(V_{2222}-V_{1111}\right) \nonumber \\ &+& N\left[\epsilon_2 - \epsilon_1-\widetilde{\Omega}(l_2-l_1)\right].
\label{calb}
\end{eqnarray} 
The semiclassical energy in the rotating frame, Eq.~(\ref{lipf}), is quadratic in $\cos \theta$ and straightforward to analyze. First of all, we notice that 
$V_{1122}<0$, because $c_0$ and $c_2$ in Eq.~(A.1) are of opposite sign, in order for the energy to be minimized. 
A minimum of $\widetilde E_{\mathrm {rot}}$ is thus 
obtained for $\phi =0$. Consequently, in the case of $f_1(\phi)$ in Eq.~(\ref{cala}), only $f_1(0)$ is here of importance.
We can furthermore identify $f_2(\widetilde{\Omega})$ as the energy difference of the system between the two many-body states $|0,N\rangle$ and $|N,0\rangle$, where all the atoms are in $|\psi_2\rangle$ and in $|\psi_1\rangle$ respectively.
At criticality, where the system passes through a correlated state of two modes of equal energy,  we thus expect $f_2(\widetilde{\Omega})$ to vanish.  
The energy Eq.~(\ref{lipf}) is 
therefore only linear in $\cos \theta$, and the value of 
$\cos \theta $ minimizing it only depends on the sign of $f_1(0)$, assuming 
$f_2(\widetilde \Omega )/f_1(0)\approx 0$.  If $f_1(0)>0$, the lowest $\widetilde{E}_{\mathrm {rot}}$ is retrieved for $\cos \theta=0$. In this case,  ${\hat J}_z \rightarrow 0$  in the semiclassical limit  and we get the ``twin" state 
\begin{eqnarray}
 |\widetilde{\Psi} \rangle = |N/2, N/2 \rangle,
\label{as1}
\end{eqnarray}
with occupation number $N/2$ for both $|\psi_1\rangle$ and $|\psi_2\rangle$. If instead $f_1(0) < 0$, then correspondingly, $\cos\theta= \pm 1$  yields the lowest energy (suggesting a superposition of $\lvert N,0 \rangle$ and $\lvert 0, N\rangle$) and we get the ``Schr\"odinger-cat"-like state
\begin{eqnarray}
 |\widetilde{\Psi} \rangle = \frac{1}{\sqrt 2} \left(|N,0\rangle + |0,N\rangle\right).
\label{as2}
\end{eqnarray}
We find that intriguingly, for large (even) atom numbers, 
there are thus only two possibilities; either the many-body ground state at criticality is given by Eq.~(\ref{as1}) or by Eq.~(\ref{as2}). 
We here stress that the reason for these two states not previously being  encountered in\,\cite{Dagnino2009a,
Dagnino2009b} lies in the relatively small number of atoms considered. 

\section{Few-body precursors of cat and twin states}

To confirm the above predictions of the two-state model and its semiclassical limit for large $N$ we now return to the exact diagonalization method. 
Clearly, our aim -- and main difficulty -- is here to search for precursors of the 
twin- and cat-like states in relation to the maximally entangled state. In general, going to larger particle numbers is a true computational challenge in the light of the increasing complexity of the quantum many-body states even in the restricted space of the lowest Landau level, further complicated by the broken symmetry due to the trap deformation. 
In practice, we first solve the single-particle problem in the rotating frame, i.e. taking into account the $\Omega \hat{L}_z$-contribution, numerically using the harmonic oscillator single-particle basis states $\phi _m $ (defined above) with $m = 0, 1, \ldots, 20$.
The full many-body state is then retrieved with a many-body basis constructed from the 
    six deformed single-particle solutions of lowest energy, which was found sufficient  around the critical frequency at the transition to the unit vortex for particle numbers up to $N=50$. 
Care is taken, as noted above, that the strength of both deformation $\alpha $ and anharmonicity $\beta $ as well as the value of $gN=1.5$ used here are compliant with using the lowest Landau level. Also, for the reasons discussed above, we only consider systems with even $N$.

\begin{figure}[H]
\includegraphics[width=\linewidth]{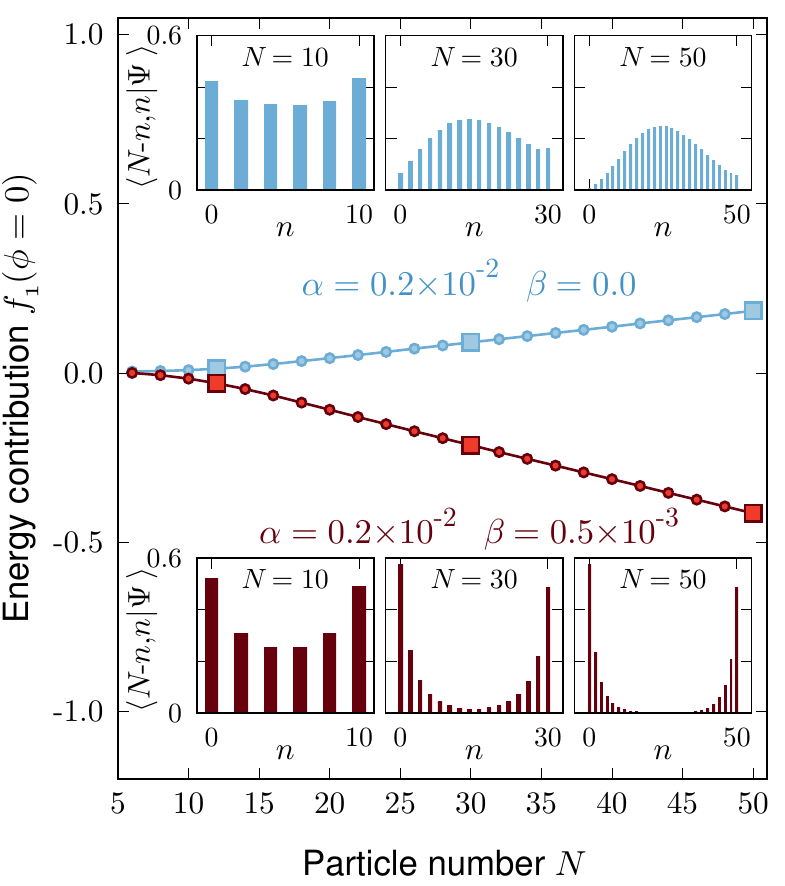}
\caption{({\it Color online.}) From maximally entangled to twin-and cat-like states for different trap settings. The quantity $f_1(0)$ from the semiclassical energy~Eq.~(\ref{lipf})  for the quadrupolar {\it (blue)} and quadrupolar plus anharmonic trap deformation {\it (red)} as specified by the corresponding blue and red labels, where $gN=1.5$ in both cases. The squares mark  the values for which the $\langle N-n,n \lvert \Psi \rangle$ distributions are shown in the top and bottom insets, for $N=10, 30$ and $50$. 
While in the limit of small $N$ the states show similarity to the maximally entangled state, in the limit of large $N$, the ground state appears as either a twin- or cat-state, depending on the form and strength of the trap deformation. 
}
\label{fig2}
\end{figure}

We here study two cases, (i) a weak quadrupolar deformation $\alpha =0.2\times  10^{-2}$ and $\beta =0$, and (ii) a weak anharmonicity in addition to (i), i.e. $\alpha =0.2\times 10^{-2}$ and $\beta =0.5\times 10^{-3}$.  For both cases we determine $\Omega _c$ by a 'regula falsi' method~\cite{Anderson1973}, with a relative error of about $10^{-5}$.
The corresponding values of $f_1(0)$, {\it i.e.}, the quantity that determines the nature of the ground state in the semiclassical limit,  
are shown as a function of the number of bosons in Fig.~\ref{fig2} for the parameters of (i)  ({\it blue}) and (ii) ({\it red}).  
We observe that the weak quadrupole deformation leads to a positive value of $f_1(0)$, and thus a distribution of the two-state coefficients favoring a twin state in the large $N$-limit, Eq.~(\ref{as1}).  Switching to  $f_1(0)<0$ can, however, be achieved by adding a weak anharmonicity as in case (ii),  where we find that  a cat-like state, Eq.~(\ref{as2}), is favored. 
As prescribed by the semiclassical analysis given above,  in both cases (i) and (ii) we find that the values of $f_1(0)$ increase roughly linearly with system size. We recall that since the value of $gN$ is fixed, $g \sim 1/N$ which, in turn, means that also $V_{ijkl}$ is approximately proportional to $1/N$ and thus that $f_1(0) \sim N$, see Eq.~(\ref{cala}).

The exact many-body ground states at criticality obtained by direct diagonalization, $\lvert \Psi (\Omega _c)\rangle $, may be characterized by their overlaps with the many-body states $\lvert N-n, n\rangle $ defined in 
Sec.~\ref{sec:CI}. 
These overlaps are shown as multiple insets in Fig.~(\ref{fig2}) at the top for (i) ({\it blue}) and bottom for (ii) ({\it red}) as a function of $n$ for $N=10, 30$ and $50$. 
Since $\lvert \Psi (\Omega _c)\rangle $ is of even parity, the overlaps are zero for odd values of $n$, as in~\cite{Dagnino2009a,Dagnino2009b}. 
For small $N$, as here shown for $N=10$, the overlaps in both cases (i) and (ii) indeed resemble the ``maximally entangled'' state of 
Refs.~\cite{Dagnino2009a,Dagnino2009b} with a distribution of next-to-equal occupancies, only slightly peaked towards maximum occupancies for $n=0$ and $n=N$. For larger $N$, however, we see a strikingly different behavior. 
In case (i), corresponding to $f_1(0) > 0$, the distribution of the overlap magnitudes does peak about $\lvert \langle N/2,N/2\lvert \Psi (\Omega _c)\rangle \rvert $. Thus, 
for these larger values of $N$, the existence of a maximally entangled state
as in~\cite{Dagnino2009a,Dagnino2009b} could not be confirmed, and instead a precursor to a twin state was found. 
An additional anharmonicity in case (ii) leads to 
$f_1(0) < 0$ where we see two equally sized peaks at 
$\lvert \langle 0,N\lvert \Psi (\Omega _c)\rangle \rvert $ and $\lvert \langle N,0\lvert \Psi (\Omega _c)\rangle \rvert $, {\it i.e.}, a cat-like distribution. 
We also observe that the twin- and cat-like distributions become more pronounced for larger $N$, as predicted by the semiclassical approach discussed in Sec.~\ref{sec3}. 

\newpage

\begin{figure}[H]
\includegraphics[width=\linewidth]{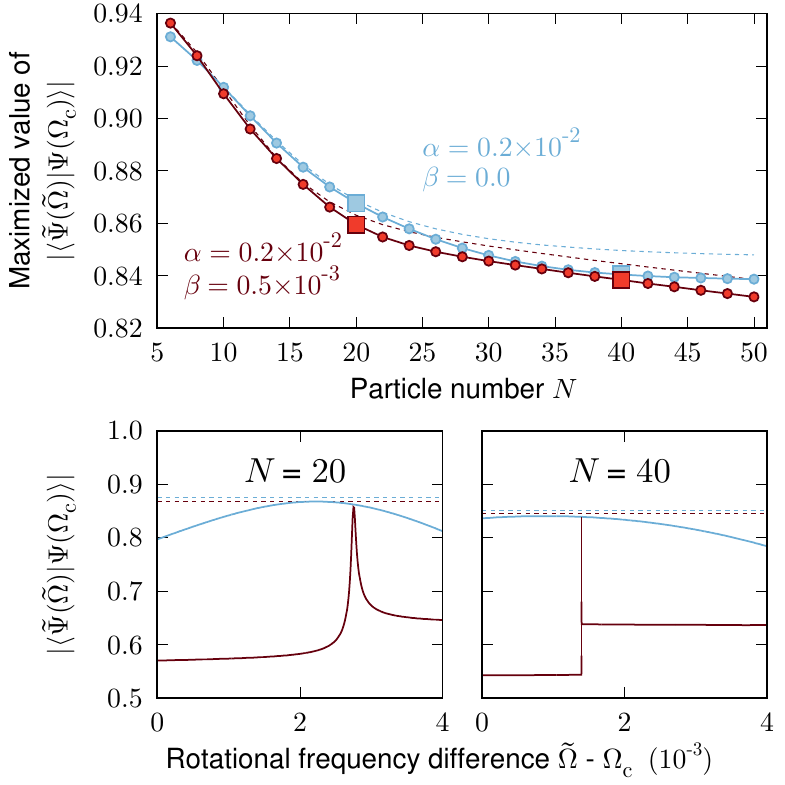}
\caption{{\it (Color online)} {\it Lower panel:} Overlap magnitudes between the full many-body state at criticality, $\lvert \Psi (\Omega _c) \rangle $, and the corresponding states $\lvert{\widetilde \Psi (\widetilde \Omega)}\rangle $, obtained in the reduced Hilbert space $\widetilde {\cal H}$ spanned by the states $\lvert N-n, n\rangle$ for $N=20$ ({\it left})  and $N=40$ ({\it right}). Here, the same trap deformations as in Fig.~\ref{fig2} are used, {\it i.e.} a quadrupolar {\it(blue line)} and a quadrupolar plus anharmonic deformation {\it(red line)}, with $\alpha$ and $\beta$ specified in the upper panel. {\it Upper panel:} The maximized magnitude of $\langle{\widetilde \Psi (\widetilde \Omega)}\lvert \Psi (\Omega _c) \rangle$, obtained at $\widetilde\Omega_c$ (identified by a numerical sweep in $\widetilde\Omega$), as a function of the particle number $N$.  In all three panels, the dashed lines indicates the upper boundary, as dictated by the overall population of the full many-body state, $\lvert \Psi(\Omega_c) \rangle$, within the subspace $\widetilde {\cal H}$.}
\label{fig3}
\end{figure}

Let us now compare the ground-state solution 
 $|\widetilde{\Psi}(\widetilde{\Omega})\rangle$
of $\hat{\widetilde{H}}_{\mathrm{rot}}$, see Eq.~(\ref{rotframe}), 
with the corresponding solution $|\Psi(\Omega_c)\rangle$ of the full
many-body Hamiltonian at criticality. The critical frequency $\widetilde{\Omega}_c$, associated with $\hat{\widetilde{H}}_{\mathrm{rot}}$, is here defined as the rotational frequency $\widetilde{\Omega}$ that maximizes the magnitude of the overlap between the two solutions $|\widetilde{\Psi}(\widetilde{\Omega})\rangle$ and $|\Psi(\Omega_c)\rangle$. In the upper panel of  Fig.~\ref{fig3},  these maxima,
{\it i.e.} $\lvert\langle\widetilde{\Psi}(\widetilde{\Omega}_c)|\Psi(\Omega_c)\rangle\rvert$, are shown for the cases (i) and (ii) considered above.
As a reference, we also include as dashed lines the
square root of the overall population of $|\Psi(\Omega_{c})\rangle$
within the reduced many-body Hilbert space constructed from the natural orbitals $|\psi_1\rangle$ and
$|\psi_2\rangle$. These latter projections set the theoretical 
upper boundary of $\lvert \langle
\widetilde{\Psi}(\widetilde{\Omega}_{c})|\Psi(\Omega_{c})\rangle\rvert $. Clearly,  the large
overlap magnitudes show that the by far largest part of the full solutions
resides within the two-state model space. Also, the fact that the
computed overlaps $\langle
\widetilde{\Psi}(\widetilde{\Omega}_{c})|\Psi(\Omega_{c})\rangle$, identified by sweeping $\widetilde \Omega $, are close
to the theoretical maxima means that the two-state model solution
captures the parts of the full solution lying in
the reduced Hilbert space $\widetilde{\cal H}$. As a consequence, for the insets of
Fig.~\ref{fig2}, we could thus as well have used $\langle N-n,
n|\widetilde{\Psi}(\widetilde{\Omega}_c)\rangle$, having a similar structure
as $\langle N-n, n|\Psi(\Omega_c)\rangle$.

Note that   $\widetilde{\Omega}_c \neq \Omega_c$, {\it i.e.}, the critical rotational frequency in the exact solution is slightly different from that of the two-state model. In the lower panel of
Fig.~\ref{fig3}, we show the overlaps obtained for $N=30$ and $N=50$
with different frequencies $\widetilde{\Omega}$ for  $\hat{\widetilde{H}}_{\mathrm{rot}}$. The lower overlaps seen for 
the rotational frequency $\widetilde{\Omega} = \Omega_c$ indicate that 
there is a subtle sensitivity of the many-body ground state to any restriction in the size of the Hilbert space right at criticality. Interestingly, however, 
the nature of the full many-body
state can largely be restored simply by using the slightly different
rotational frequency $\widetilde{\Omega}_{c}$ for the two-state 
Hamiltonian (without changing $|\psi_1\rangle$, $|\psi_2\rangle$ and
$g$). Also, although not shown, $f_2\approx 0$ when $\widetilde{\Omega} = \widetilde{\Omega}_c$.

\begin{figure}[H]
\includegraphics[width=0.9\linewidth]{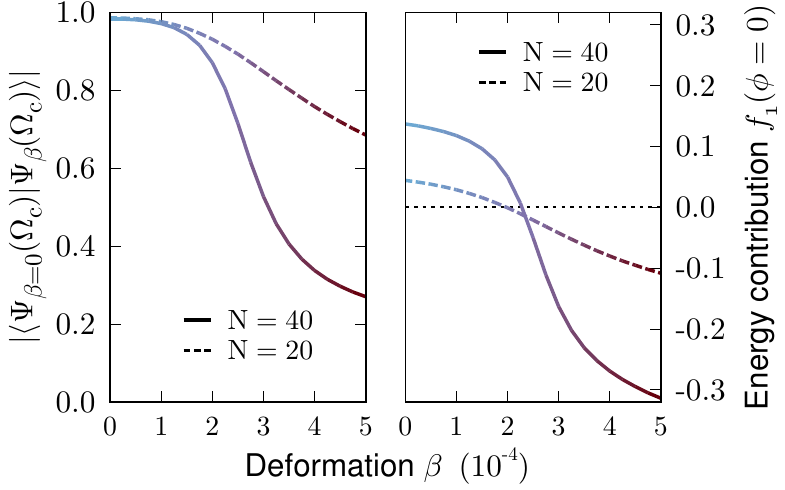}
\caption{{\it (Color online)} The robustness of the many-body state at criticality for a change in $\beta$ (when $\alpha = 0.2\times 10^{-2}$ and $gN=1.5$). The left panel shows the magnitude of the overlap between the many-body state at a given $\beta$ and the corresponding one at $\beta=0$, {\it i.e.}, $\lvert \langle \Psi_{\beta=0}(\Omega_c)\lvert \Psi_{\beta}(\Omega_c)\rangle \rvert $, for $N=20$ and $N=40$. The sudden drop in the overlap, observed for $N=40$, occurs when $f_1(0)$ changes sign, see  right panel. (The dashed line is a guide to the eye).}
\label{fig4}
\end{figure}

Finally, we stress that the considered natural orbitals $\lvert \psi_1\rangle$ and $\lvert \psi_2\rangle$  generally  depend on the shape of the trap as well as on the particle number. Hence, when the system goes from a twin-like distribution in case (i) to a cat-like one in case (ii) it is not transparent to what degree this change in distribution reflects an actual change in the many-body state $\lvert \Psi(\Omega_c)\rangle$. In the left panel of Fig.~\ref{fig4}, we show the overlap between the state $\lvert {\Psi}(\Omega_c) \rangle $ obtained for case (i) and the corresponding state obtained with an additional anharmonic deformation of strength $\beta$, for $N=20$ and $N=40$. For $N=20$ a smooth decrease in overlap can be seen. For the larger $N$, however, a sharper transition at $\beta \sim 2 \times 10^{-4}$ is observed. Also, the swift change in $\lvert \Psi(\Omega_c)\rangle$ occurs exactly where $f_1(0)$ changes sign, see the right panel. 
We may thus conclude that 
for large $N$ and for $\beta $ up to a certain value, the many-body state (described by a twin-like distribution) stays fairly much the same. If we increase $\beta$ beyond this point,  the many-body state changes its  structure, with a distribution of occupancies resembling that of a cat-like state. In practice, to reduce the computational cost, we here only account for the part of $|\Psi(\Omega_c)\rangle$ that resides in the reduced Hilbert space spanned by the four natural orbitals with largest occupancy ($\psi_1, \psi_2, \psi_3$ and $\psi_4$) with the additional constraint that $n_3 + n_4 \leq 4$, where $n_i$ is the occupation number of $\psi_i$. We do thus include, and go beyond, the space spanned by the states $\lvert N-n,n\rangle$ (which, as discussed above, already covers most of $\lvert \Psi(\Omega_c)\rangle$). 
In fact,  the obtained norm of $\lvert \Psi (\Omega _c)\rangle $ is with this approach  always $> 0.99$ for the considered 
values of $\beta$ and $N$. The limited Hilbert space thus seems adequate, justifying our conclusion that a change in the $\langle N-n,n \lvert \Psi(\Omega_c) \rangle$ distibution also reflects a change in the many-body state $\lvert \Psi(\Omega_c)\rangle$. In addition,  we find that the transition between a twin- and cat-like state becomes more abrupt for larger $N$.

\section{Conclusions}

The vortex nucleation process in a rotating scalar Bose-Einstein condensate provides a unique model system to study the emergence of a quantum phase transition from the microscopic few-body regime to the thermodynamic limit. 
Even in the presence of interactions, 
the nucleation process of the first vortex is associated with exact linearity of the ground state energy as a function of angular momentum  which leads to a discontinuous transition between the non-rotating ground state  and the unit vortex. This peculiarity makes the
nucleation of the first vortex a particularly interesting scenario to study the 
nature of the phase transition,  enabling a direct comparison between a next-to-exact numerical approach and the Gross-Pitaevskii mean-field solution. 

It was earlier found that the formation of the first vortex localized at the center of the rotating cloud passes through a quantum critical point, where 
two of the macroscopically occupied natural orbitals have equal occupancy~\cite{Dagnino2009a,Dagnino2009b}. It was pointed out that in the limit of small $N$ and sufficiently weak interactions, the transition gives rise to a maximally entangled many-body state that invalidates the Gross-Pitaevskii approach when passing through criticality. 

Here, we developed a two-state model similar to the 
Lipkin-Meshkov-Glick model~\cite{Lipkin1965}, also following its adaption to Bose gases in Ref.~\cite{Elgaroy1999}.  
We found that for a weak quadrupolar deformation of the trap, 
the maximally entangled state at criticality prevails for small particle numbers. In the large-$N$ limit, however, the states rather resemble ``cat"- or ``twin"-like states, depending on the perturbative shape of the confinement, being quadrupolar or also with an added quartic contribution (as seen in Fig.~\ref{fig2} which summarizes the main result of this paper).  We corroborated the validity of the LMG model by numerical exact diagonalization in the lowest Landau level for sizes $N\lesssim 50$. The larger $N$, the more abrupt  this transition becomes. 
 
From our analysis it became obvious that an experimental realization of these
correlated states at criticality would need a fine-tuning of trap deformation,  rotational frequencies as well as particle numbers (being even or odd) that 
is next-to impossible to achieve. The conditions for their realization appear most favorable in the limit of moderate atom numbers of just a few dozen where finite-size effects still prevail.
The value of the present study thus mainly lies in the study of the transition through quantum criticality from the few- to the many-particle regime.

In future work, it would be interesting to try to extract the exact nature of the nucleation of the first vortex  in a deformed trap from an analysis of the exact ground state wave function up to the unit vortex. The latter was analytically derived for the case of azimuthal trap symmetry~\cite{Bertsch1999,Jackson2000,Papenbrock2001}.  The unit vortex resembles one of the very few examples where the exact many-body ground state is known analytically. Perturbatively extracting the twin- and cat-like states discussed here from this exact many-body state opens an intriguing (yet difficult) way to analytically address the vortex nucleation process which however goes beyond the scope of this work. 

\acknowledgements

This work was financially supported by the Swedish Research Council and the Knut and Alice Wallenberg foundation. 

\appendix*

\section{Mean-field results for axially broken symmetry}

When the symmetry-breaking quadrupolar potential 
$\Delta V=\alpha (x^2-y^2)$ is weak, {\it i.e.}, the 
associated shift in energy is much smaller than the interaction energy (which in turn is also much smaller than the oscillator quantum of energy) we may apply perturbation theory. 
In Ref.\,\cite{Kavoulakis2000} we have seen earlier that in the absence of $\Delta V$ and for $\ell \to 0^+$, where 
$\ell = L/N$, the order parameter $\psi ^{(\ell )}$ has the form 
\begin{equation}
\psi ^{(\ell )} = \psi ^{(0^+)} \approx c_0 \phi_0 + c_2 \phi_2,
 \label{2v}
\end{equation}
where
\begin{equation}
  \phi_m = \frac 1 {\sqrt{\pi m!}} r^m e^{i m \vartheta} e^{- r^2/2} 
  \label{spfunction}
\end{equation}
and $|c_0|^2 = 1 - \ell/2$, $|c_2|^2 = \ell/2$.

Evaluating the expectation value  $\langle \Delta V\rangle $ in $\psi ^{(0^+)}$ 
we find that $ \langle \Delta V \rangle = \sqrt 2 \, \alpha  c_0 c_2 = \frac 1 {\sqrt 2} \alpha  \sqrt{\ell (2 - \ell)}$ and 
thus for the energy per particle for $\ell \to 0^+$ 
\begin{equation}
   E(\ell) =  \frac {N g} {4 \pi} + \ell \left(1 - \Omega - \frac {N g} {8 \pi} \right) 
   - \frac  {\alpha  \sqrt{\ell (2 - \ell) }}{\sqrt 2}.
   \label{en1}
\end{equation}
It is interesting to note that for $\ell \to 0^+$, $\Delta V$ gives a term which scales as $\sqrt \ell$. For 
$\ell \to 0^+$ the angular momentum is carried by the $m=2$ state, or in other words, with the small quadrupolar deformation, there are now two vortices 
entering the cloud (from opposite sides) from infinity with increasing rotation. 
The potential $\Delta V$ has a two-fold symmetry and 
thus the dominant $m=0$ state is coupled via $\Delta V$ with the $m=2$ state. This is the reason why $\langle 
\Delta V \rangle$ scales as $c_0 c_2 \propto \sqrt \ell$, for $\ell \to 0$. 

The opposite limit, $\ell \to 1^-$, is more tricky. We recall~\cite{Kavoulakis2000} that when $\Delta V = 0$ the order parameter is 
\begin{equation}
 \psi ^{(1^-)} \approx c_0 \phi_0 + c_1 \phi_1 + c_2 \phi_2,
\end{equation}
where $|c_0|^2 = 2 (1 - \ell)$, $|c_1|^2 = 1 - 3 (1-\ell)$, and $|c_2|^2 = 1-\ell$. While one may be tempted to perform the
same calculation as before, this would not be quite correct. The reason is that for $\ell \approx 1^-$ 
the state $\phi_3$ has a non-negligible contribution to the order parameter. This is not a surprise, since $\phi_3$ can couple with $\phi_1$ via $\Delta V$ and thus lower the energy. 
To see the effect of the $\phi_3$ state, let us focus at the value of $\ell$ where only $c_1$ and $c_3$ are
nonzero (for a value of $\ell$ somewhat larger than unity). Considering the order parameter
\begin{equation}
 \psi ^{(1^+)} = c_1 \phi_1 + c_3 \phi_3,
 \label{3v}
\end{equation}
and minimizing the energy under the constraints $|c_1|^2 + |c_3|^2 = 1$ and $|c_1|^2 + 3 |c_3|^2 = \ell$, we obtain
\begin{equation}
|c_1|^2 = \frac 1 2 (3 - \ell), \,\,\, |c_3|^2 = \frac 1 2 (\ell - 1).
\end{equation} 
Setting $\ell = 1 + \epsilon$, with $\epsilon$ being small,
\begin{equation}
|c_1|^2 = 1 - \frac {\epsilon} 2, \,\,\, |c_3|^2 = \frac {\epsilon} 2.
\end{equation}   
The corresponding energy per particle in the rotating frame is 
\begin{eqnarray}
 \frac E N &=&  \ell (1 - \Omega) 
 + {\small \frac {g (N-1)} {2 \pi}
  \left( \frac {|c_1|^4} 4 + \frac {5 |c_3|^4} {32} +
 \frac { |c_1|^2 |c_3|^2} 2 \right)} \nonumber \\
  &-& {\sqrt 6} \alpha c_1 c_3.
\end{eqnarray}
Expanding in $\epsilon$ we obtain
\begin{equation}
\frac E N = (1-\Omega) (1+\epsilon) - \alpha \sqrt{3 \epsilon} + {\cal O}(\epsilon^2).
\end{equation}
It is interesting that there is no linear term in $\epsilon$ that comes from the interaction, which,
however enters via $\Omega$, as we see below. Minimizing the energy we find that 
\begin{equation}
  \epsilon_0 = \frac {3 \alpha ^2} {4 (1-\Omega)^2}
\end{equation}
and thus the corresponding value of $\ell$ is 
\begin{equation}
 \ell_0 = 1 + \frac {3 \alpha ^2} {4 (1-\Omega)^2}.
\end{equation} 
Using for $\Omega$ the critical value $\tilde{\Omega}_c = 1 - gN/(8 \pi)$, then
\begin{equation}
 \ell_0 = 1 + 48 \pi^2 \frac {\alpha ^2} {(N g)^2}.
\label{res1}
\end{equation} 
The obtained correction is of order $[\alpha /(gN)]^2$, which is $\ll 1$. The corresponding values of $c_1$ and $c_3$ are
\begin{eqnarray}
|c_1|^2 &= 1 - 24 \pi^2 \frac {\alpha ^2} {(N g)^2}, \nonumber \\ 
|c_3|^2 &= 24 \pi^2 \frac {\alpha ^2} {(N g)^2}, \nonumber
\label{exp}
\end{eqnarray} 
while the corresponding energy per particle is 
\begin{equation}
{E \over N} = {gN \over 8 \pi} \left( 1 - 48 \pi^2  {\alpha ^2\over (N g)^2} \right) ~.
\end{equation}
\\
Therefore, $\Delta V$ shifts the value of $\ell$ where $|c_1|^2$ takes its maximum value from unity to a slightly 
larger value. This is a single-particle effect and the interaction does not play any role here.  We thus observe that while in the limit $\ell \to 0^+$ the order parameter is a linear superposition of $\phi_0$ 
and $\phi_2$ both due to the interaction and due to $\Delta V$, in the limit $\ell \to 1^-$ this is not the case. As a result, for $\ell \to 0^+$ the parameter $\alpha $ appears linearly, but for  $\ell \to 1^-$, it appears quadratically in the energy. 

To summarize the above, we see that within the mean field approximation for values of $\ell $ close to zero the order parameter is very well approximated by Eq.~\ref{2v}. For $\ell \rightarrow 1^-$ 
the order parameter is approximately equal to $\phi _1$ plus corrections of order $\alpha ^2 /(gN)^2$.


\begin{thebibliography}{72}%
\makeatletter
\providecommand \@ifxundefined [1]{%
 \@ifx{#1\undefined}
}%
\providecommand \@ifnum [1]{%
 \ifnum #1\expandafter \@firstoftwo
 \else \expandafter \@secondoftwo
 \fi
}%
\providecommand \@ifx [1]{%
 \ifx #1\expandafter \@firstoftwo
 \else \expandafter \@secondoftwo
 \fi
}%
\providecommand \natexlab [1]{#1}%
\providecommand \enquote  [1]{``#1''}%
\providecommand \bibnamefont  [1]{#1}%
\providecommand \bibfnamefont [1]{#1}%
\providecommand \citenamefont [1]{#1}%
\providecommand \href@noop [0]{\@secondoftwo}%
\providecommand \href [0]{\begingroup \@sanitize@url \@href}%
\providecommand \@href[1]{\@@startlink{#1}\@@href}%
\providecommand \@@href[1]{\endgroup#1\@@endlink}%
\providecommand \@sanitize@url [0]{\catcode `\\12\catcode `\$12\catcode
  `\&12\catcode `\#12\catcode `\^12\catcode `\_12\catcode `\%12\relax}%
\providecommand \@@startlink[1]{}%
\providecommand \@@endlink[0]{}%
\providecommand \url  [0]{\begingroup\@sanitize@url \@url }%
\providecommand \@url [1]{\endgroup\@href {#1}{\urlprefix }}%
\providecommand \urlprefix  [0]{URL }%
\providecommand \Eprint [0]{\href }%
\providecommand \doibase [0]{http://dx.doi.org/}%
\providecommand \selectlanguage [0]{\@gobble}%
\providecommand \bibinfo  [0]{\@secondoftwo}%
\providecommand \bibfield  [0]{\@secondoftwo}%
\providecommand \translation [1]{[#1]}%
\providecommand \BibitemOpen [0]{}%
\providecommand \bibitemStop [0]{}%
\providecommand \bibitemNoStop [0]{.\EOS\space}%
\providecommand \EOS [0]{\spacefactor3000\relax}%
\providecommand \BibitemShut  [1]{\csname bibitem#1\endcsname}%
\let\auto@bib@innerbib\@empty
\bibitem [{\citenamefont {Dagnino}\ \emph
  {et~al.}(2009{\natexlab{a}})\citenamefont {Dagnino}, \citenamefont
  {Barber{\'a}n}, \citenamefont {Lewenstein},\ and\ \citenamefont
  {Dalibard}}]{Dagnino2009a}%
  \BibitemOpen
  \bibfield  {author} {\bibinfo {author} {\bibfnamefont {D.}~\bibnamefont
  {Dagnino}}, \bibinfo {author} {\bibfnamefont {N.}~\bibnamefont
  {Barber{\'a}n}}, \bibinfo {author} {\bibfnamefont {M.}~\bibnamefont
  {Lewenstein}}, \ and\ \bibinfo {author} {\bibfnamefont {J.}~\bibnamefont
  {Dalibard}},\ }\href {https://doi.org/10.1038/nphys1277} {\bibfield
  {journal} {\bibinfo  {journal} {Nature Physics}\ }\textbf {\bibinfo {volume}
  {5}},\ \bibinfo {pages} {431 EP } (\bibinfo {year}
  {2009}{\natexlab{a}})}\BibitemShut {NoStop}%
\bibitem [{\citenamefont {Pethick}\ and\ \citenamefont
  {Smith}(2008)}]{Pethick}%
  \BibitemOpen
  \bibfield  {author} {\bibinfo {author} {\bibfnamefont {C.}~\bibnamefont
  {Pethick}}\ and\ \bibinfo {author} {\bibfnamefont {H.}~\bibnamefont
  {Smith}},\ }\href {\doibase 10.1017/CBO9780511802850} {\emph {\bibinfo
  {title} {Bose - Einstein // condensation in dilute gases}}},\ \bibinfo
  {edition} {2nd}\ ed.\ (\bibinfo  {publisher} {Cambridge University Press},\
  \bibinfo {year} {2008})\BibitemShut {NoStop}%
\bibitem [{\citenamefont {P{\'\i}tajevsk{\'\i}j}\ and\ \citenamefont
  {Stringari}(2003)}]{Stringari}%
  \BibitemOpen
  \bibfield  {author} {\bibinfo {author} {\bibfnamefont {L.}~\bibnamefont
  {P{\'\i}tajevsk{\'\i}j}}\ and\ \bibinfo {author} {\bibfnamefont
  {S.}~\bibnamefont {Stringari}},\ }\href
  {https://books.google.de/books?id=rIobbOxC4j4C} {\emph {\bibinfo {title}
  {Bose-Einstein Condensation}}}\ (\bibinfo  {publisher} {Clarendon Press},\
  \bibinfo {year} {2003})\BibitemShut {NoStop}%
\bibitem [{\citenamefont {Chevy}\ \emph {et~al.}(2000)\citenamefont {Chevy},
  \citenamefont {Madison},\ and\ \citenamefont {Dalibard}}]{Chevy2000}%
  \BibitemOpen
  \bibfield  {author} {\bibinfo {author} {\bibfnamefont {F.}~\bibnamefont
  {Chevy}}, \bibinfo {author} {\bibfnamefont {K.~W.}\ \bibnamefont {Madison}},
  \ and\ \bibinfo {author} {\bibfnamefont {J.}~\bibnamefont {Dalibard}},\
  }\href {\doibase 10.1103/PhysRevLett.85.2223} {\bibfield  {journal} {\bibinfo
   {journal} {Phys. Rev. Lett.}\ }\textbf {\bibinfo {volume} {85}},\ \bibinfo
  {pages} {2223} (\bibinfo {year} {2000})}\BibitemShut {NoStop}%
\bibitem [{\citenamefont {Madison}\ \emph {et~al.}(2000)\citenamefont
  {Madison}, \citenamefont {Chevy}, \citenamefont {Wohlleben},\ and\
  \citenamefont {Dalibard}}]{Madison2000}%
  \BibitemOpen
  \bibfield  {author} {\bibinfo {author} {\bibfnamefont {K.~W.}\ \bibnamefont
  {Madison}}, \bibinfo {author} {\bibfnamefont {F.}~\bibnamefont {Chevy}},
  \bibinfo {author} {\bibfnamefont {W.}~\bibnamefont {Wohlleben}}, \ and\
  \bibinfo {author} {\bibfnamefont {J.}~\bibnamefont {Dalibard}},\ }\href
  {\doibase 10.1103/PhysRevLett.84.806} {\bibfield  {journal} {\bibinfo
  {journal} {Phys. Rev. Lett.}\ }\textbf {\bibinfo {volume} {84}},\ \bibinfo
  {pages} {806} (\bibinfo {year} {2000})}\BibitemShut {NoStop}%
\bibitem [{\citenamefont {Madison}\ \emph {et~al.}(2001)\citenamefont
  {Madison}, \citenamefont {Chevy}, \citenamefont {Bretin},\ and\ \citenamefont
  {Dalibard}}]{Madison2001}%
  \BibitemOpen
  \bibfield  {author} {\bibinfo {author} {\bibfnamefont {K.~W.}\ \bibnamefont
  {Madison}}, \bibinfo {author} {\bibfnamefont {F.}~\bibnamefont {Chevy}},
  \bibinfo {author} {\bibfnamefont {V.}~\bibnamefont {Bretin}}, \ and\ \bibinfo
  {author} {\bibfnamefont {J.}~\bibnamefont {Dalibard}},\ }\href {\doibase
  10.1103/PhysRevLett.86.4443} {\bibfield  {journal} {\bibinfo  {journal}
  {Phys. Rev. Lett.}\ }\textbf {\bibinfo {volume} {86}},\ \bibinfo {pages}
  {4443} (\bibinfo {year} {2001})}\BibitemShut {NoStop}%
\bibitem [{\citenamefont {Haljan}\ \emph {et~al.}(2001)\citenamefont {Haljan},
  \citenamefont {Coddington}, \citenamefont {Engels},\ and\ \citenamefont
  {Cornell}}]{Haljan2001}%
  \BibitemOpen
  \bibfield  {author} {\bibinfo {author} {\bibfnamefont {P.~C.}\ \bibnamefont
  {Haljan}}, \bibinfo {author} {\bibfnamefont {I.}~\bibnamefont {Coddington}},
  \bibinfo {author} {\bibfnamefont {P.}~\bibnamefont {Engels}}, \ and\ \bibinfo
  {author} {\bibfnamefont {E.~A.}\ \bibnamefont {Cornell}},\ }\href {\doibase
  10.1103/PhysRevLett.87.210403} {\bibfield  {journal} {\bibinfo  {journal}
  {Phys. Rev. Lett.}\ }\textbf {\bibinfo {volume} {87}},\ \bibinfo {pages}
  {210403} (\bibinfo {year} {2001})}\BibitemShut {NoStop}%
\bibitem [{\citenamefont {Hodby}\ \emph {et~al.}(2001)\citenamefont {Hodby},
  \citenamefont {Hechenblaikner}, \citenamefont {Hopkins}, \citenamefont
  {Marag\`o},\ and\ \citenamefont {Foot}}]{Hodby2001}%
  \BibitemOpen
  \bibfield  {author} {\bibinfo {author} {\bibfnamefont {E.}~\bibnamefont
  {Hodby}}, \bibinfo {author} {\bibfnamefont {G.}~\bibnamefont
  {Hechenblaikner}}, \bibinfo {author} {\bibfnamefont {S.~A.}\ \bibnamefont
  {Hopkins}}, \bibinfo {author} {\bibfnamefont {O.~M.}\ \bibnamefont
  {Marag\`o}}, \ and\ \bibinfo {author} {\bibfnamefont {C.~J.}\ \bibnamefont
  {Foot}},\ }\href {\doibase 10.1103/PhysRevLett.88.010405} {\bibfield
  {journal} {\bibinfo  {journal} {Phys. Rev. Lett.}\ }\textbf {\bibinfo
  {volume} {88}},\ \bibinfo {pages} {010405} (\bibinfo {year}
  {2001})}\BibitemShut {NoStop}%
\bibitem [{\citenamefont {Raman}\ \emph {et~al.}(2001)\citenamefont {Raman},
  \citenamefont {Abo-Shaeer}, \citenamefont {Vogels}, \citenamefont {Xu},\ and\
  \citenamefont {Ketterle}}]{Raman2001}%
  \BibitemOpen
  \bibfield  {author} {\bibinfo {author} {\bibfnamefont {C.}~\bibnamefont
  {Raman}}, \bibinfo {author} {\bibfnamefont {J.~R.}\ \bibnamefont
  {Abo-Shaeer}}, \bibinfo {author} {\bibfnamefont {J.~M.}\ \bibnamefont
  {Vogels}}, \bibinfo {author} {\bibfnamefont {K.}~\bibnamefont {Xu}}, \ and\
  \bibinfo {author} {\bibfnamefont {W.}~\bibnamefont {Ketterle}},\ }\href
  {\doibase 10.1103/PhysRevLett.87.210402} {\bibfield  {journal} {\bibinfo
  {journal} {Phys. Rev. Lett.}\ }\textbf {\bibinfo {volume} {87}},\ \bibinfo
  {pages} {210402} (\bibinfo {year} {2001})}\BibitemShut {NoStop}%
\bibitem [{\citenamefont {Abo-Shaeer}\ \emph {et~al.}(2001)\citenamefont
  {Abo-Shaeer}, \citenamefont {Raman}, \citenamefont {Vogels},\ and\
  \citenamefont {Ketterle}}]{Abo-Shaeer2001}%
  \BibitemOpen
  \bibfield  {author} {\bibinfo {author} {\bibfnamefont {J.~R.}\ \bibnamefont
  {Abo-Shaeer}}, \bibinfo {author} {\bibfnamefont {C.}~\bibnamefont {Raman}},
  \bibinfo {author} {\bibfnamefont {J.~M.}\ \bibnamefont {Vogels}}, \ and\
  \bibinfo {author} {\bibfnamefont {W.}~\bibnamefont {Ketterle}},\ }\href
  {\doibase 10.1126/science.1060182} {\bibfield  {journal} {\bibinfo  {journal}
  {Science}\ }\textbf {\bibinfo {volume} {292}},\ \bibinfo {pages} {476}
  (\bibinfo {year} {2001})}\BibitemShut {NoStop}%
\bibitem [{\citenamefont {Abo-Shaeer}\ \emph {et~al.}(2002)\citenamefont
  {Abo-Shaeer}, \citenamefont {Raman},\ and\ \citenamefont
  {Ketterle}}]{Abo-Shaeer2002}%
  \BibitemOpen
  \bibfield  {author} {\bibinfo {author} {\bibfnamefont {J.~R.}\ \bibnamefont
  {Abo-Shaeer}}, \bibinfo {author} {\bibfnamefont {C.}~\bibnamefont {Raman}}, \
  and\ \bibinfo {author} {\bibfnamefont {W.}~\bibnamefont {Ketterle}},\ }\href
  {\doibase 10.1103/PhysRevLett.88.070409} {\bibfield  {journal} {\bibinfo
  {journal} {Phys. Rev. Lett.}\ }\textbf {\bibinfo {volume} {88}},\ \bibinfo
  {pages} {070409} (\bibinfo {year} {2002})}\BibitemShut {NoStop}%
\bibitem [{\citenamefont {Engels}\ \emph {et~al.}(2002)\citenamefont {Engels},
  \citenamefont {Coddington}, \citenamefont {Haljan},\ and\ \citenamefont
  {Cornell}}]{Engels2002}%
  \BibitemOpen
  \bibfield  {author} {\bibinfo {author} {\bibfnamefont {P.}~\bibnamefont
  {Engels}}, \bibinfo {author} {\bibfnamefont {I.}~\bibnamefont {Coddington}},
  \bibinfo {author} {\bibfnamefont {P.~C.}\ \bibnamefont {Haljan}}, \ and\
  \bibinfo {author} {\bibfnamefont {E.~A.}\ \bibnamefont {Cornell}},\ }\href
  {\doibase 10.1103/PhysRevLett.89.100403} {\bibfield  {journal} {\bibinfo
  {journal} {Phys. Rev. Lett.}\ }\textbf {\bibinfo {volume} {89}},\ \bibinfo
  {pages} {100403} (\bibinfo {year} {2002})}\BibitemShut {NoStop}%
\bibitem [{\citenamefont {Engels}\ \emph {et~al.}(2003)\citenamefont {Engels},
  \citenamefont {Coddington}, \citenamefont {Haljan}, \citenamefont
  {Schweikhard},\ and\ \citenamefont {Cornell}}]{Engels2003}%
  \BibitemOpen
  \bibfield  {author} {\bibinfo {author} {\bibfnamefont {P.}~\bibnamefont
  {Engels}}, \bibinfo {author} {\bibfnamefont {I.}~\bibnamefont {Coddington}},
  \bibinfo {author} {\bibfnamefont {P.~C.}\ \bibnamefont {Haljan}}, \bibinfo
  {author} {\bibfnamefont {V.}~\bibnamefont {Schweikhard}}, \ and\ \bibinfo
  {author} {\bibfnamefont {E.~A.}\ \bibnamefont {Cornell}},\ }\href {\doibase
  10.1103/PhysRevLett.90.170405} {\bibfield  {journal} {\bibinfo  {journal}
  {Phys. Rev. Lett.}\ }\textbf {\bibinfo {volume} {90}},\ \bibinfo {pages}
  {170405} (\bibinfo {year} {2003})}\BibitemShut {NoStop}%
\bibitem [{\citenamefont {Schweikhard}\ \emph {et~al.}(2004)\citenamefont
  {Schweikhard}, \citenamefont {Coddington}, \citenamefont {Engels},
  \citenamefont {Mogendorff},\ and\ \citenamefont {Cornell}}]{Schweikhard2004}%
  \BibitemOpen
  \bibfield  {author} {\bibinfo {author} {\bibfnamefont {V.}~\bibnamefont
  {Schweikhard}}, \bibinfo {author} {\bibfnamefont {I.}~\bibnamefont
  {Coddington}}, \bibinfo {author} {\bibfnamefont {P.}~\bibnamefont {Engels}},
  \bibinfo {author} {\bibfnamefont {V.~P.}\ \bibnamefont {Mogendorff}}, \ and\
  \bibinfo {author} {\bibfnamefont {E.~A.}\ \bibnamefont {Cornell}},\ }\href
  {\doibase 10.1103/PhysRevLett.92.040404} {\bibfield  {journal} {\bibinfo
  {journal} {Phys. Rev. Lett.}\ }\textbf {\bibinfo {volume} {92}},\ \bibinfo
  {pages} {040404} (\bibinfo {year} {2004})}\BibitemShut {NoStop}%
\bibitem [{\citenamefont {Barranco}\ \emph {et~al.}(2006)\citenamefont
  {Barranco}, \citenamefont {Guardiola}, \citenamefont {Hernández},
  \citenamefont {Mayol}, \citenamefont {Navarro},\ and\ \citenamefont
  {Pi}}]{Barranco2006}%
  \BibitemOpen
  \bibfield  {author} {\bibinfo {author} {\bibfnamefont {M.}~\bibnamefont
  {Barranco}}, \bibinfo {author} {\bibfnamefont {R.}~\bibnamefont {Guardiola}},
  \bibinfo {author} {\bibfnamefont {S.}~\bibnamefont {Hernández}}, \bibinfo
  {author} {\bibfnamefont {R.}~\bibnamefont {Mayol}}, \bibinfo {author}
  {\bibfnamefont {J.}~\bibnamefont {Navarro}}, \ and\ \bibinfo {author}
  {\bibfnamefont {M.}~\bibnamefont {Pi}},\ }\href {\doibase
  10.1007/s10909-005-9267-0} {\bibfield  {journal} {\bibinfo  {journal}
  {Journal of Low Temperature Physics}\ }\textbf {\bibinfo {volume} {142}},\
  \bibinfo {pages} {1} (\bibinfo {year} {2006})}\BibitemShut {NoStop}%
\bibitem [{\citenamefont {Gomez}\ \emph {et~al.}(2014)\citenamefont {Gomez},
  \citenamefont {Ferguson}, \citenamefont {Cryan}, \citenamefont {Bacellar},
  \citenamefont {Tanyag}, \citenamefont {Jones}, \citenamefont {Schorb},
  \citenamefont {Anielski}, \citenamefont {Belkacem}, \citenamefont {Bernando},
  \citenamefont {Boll}, \citenamefont {Bozek}, \citenamefont {Carron},
  \citenamefont {Chen}, \citenamefont {Delmas}, \citenamefont {Englert},
  \citenamefont {Epp}, \citenamefont {Erk}, \citenamefont {Foucar},
  \citenamefont {Hartmann}, \citenamefont {Hexemer}, \citenamefont {Huth},
  \citenamefont {Kwok}, \citenamefont {Leone}, \citenamefont {Ma},
  \citenamefont {Maia}, \citenamefont {Malmerberg}, \citenamefont {Marchesini},
  \citenamefont {Neumark}, \citenamefont {Poon}, \citenamefont {Prell},
  \citenamefont {Rolles}, \citenamefont {Rudek}, \citenamefont {Rudenko},
  \citenamefont {Seifrid}, \citenamefont {Siefermann}, \citenamefont {Sturm},
  \citenamefont {Swiggers}, \citenamefont {Ullrich}, \citenamefont {Weise},
  \citenamefont {Zwart}, \citenamefont {Bostedt}, \citenamefont {Gessner},\
  and\ \citenamefont {Vilesov}}]{Gomez2014}%
  \BibitemOpen
  \bibfield  {author} {\bibinfo {author} {\bibfnamefont {L.~F.}\ \bibnamefont
  {Gomez}}, \bibinfo {author} {\bibfnamefont {K.~R.}\ \bibnamefont {Ferguson}},
  \bibinfo {author} {\bibfnamefont {J.~P.}\ \bibnamefont {Cryan}}, \bibinfo
  {author} {\bibfnamefont {C.}~\bibnamefont {Bacellar}}, \bibinfo {author}
  {\bibfnamefont {R.~M.~P.}\ \bibnamefont {Tanyag}}, \bibinfo {author}
  {\bibfnamefont {C.}~\bibnamefont {Jones}}, \bibinfo {author} {\bibfnamefont
  {S.}~\bibnamefont {Schorb}}, \bibinfo {author} {\bibfnamefont
  {D.}~\bibnamefont {Anielski}}, \bibinfo {author} {\bibfnamefont
  {A.}~\bibnamefont {Belkacem}}, \bibinfo {author} {\bibfnamefont
  {C.}~\bibnamefont {Bernando}}, \bibinfo {author} {\bibfnamefont
  {R.}~\bibnamefont {Boll}}, \bibinfo {author} {\bibfnamefont {J.}~\bibnamefont
  {Bozek}}, \bibinfo {author} {\bibfnamefont {S.}~\bibnamefont {Carron}},
  \bibinfo {author} {\bibfnamefont {G.}~\bibnamefont {Chen}}, \bibinfo {author}
  {\bibfnamefont {T.}~\bibnamefont {Delmas}}, \bibinfo {author} {\bibfnamefont
  {L.}~\bibnamefont {Englert}}, \bibinfo {author} {\bibfnamefont {S.~W.}\
  \bibnamefont {Epp}}, \bibinfo {author} {\bibfnamefont {B.}~\bibnamefont
  {Erk}}, \bibinfo {author} {\bibfnamefont {L.}~\bibnamefont {Foucar}},
  \bibinfo {author} {\bibfnamefont {R.}~\bibnamefont {Hartmann}}, \bibinfo
  {author} {\bibfnamefont {A.}~\bibnamefont {Hexemer}}, \bibinfo {author}
  {\bibfnamefont {M.}~\bibnamefont {Huth}}, \bibinfo {author} {\bibfnamefont
  {J.}~\bibnamefont {Kwok}}, \bibinfo {author} {\bibfnamefont {S.~R.}\
  \bibnamefont {Leone}}, \bibinfo {author} {\bibfnamefont {J.~H.~S.}\
  \bibnamefont {Ma}}, \bibinfo {author} {\bibfnamefont {F.~R. N.~C.}\
  \bibnamefont {Maia}}, \bibinfo {author} {\bibfnamefont {E.}~\bibnamefont
  {Malmerberg}}, \bibinfo {author} {\bibfnamefont {S.}~\bibnamefont
  {Marchesini}}, \bibinfo {author} {\bibfnamefont {D.~M.}\ \bibnamefont
  {Neumark}}, \bibinfo {author} {\bibfnamefont {B.}~\bibnamefont {Poon}},
  \bibinfo {author} {\bibfnamefont {J.}~\bibnamefont {Prell}}, \bibinfo
  {author} {\bibfnamefont {D.}~\bibnamefont {Rolles}}, \bibinfo {author}
  {\bibfnamefont {B.}~\bibnamefont {Rudek}}, \bibinfo {author} {\bibfnamefont
  {A.}~\bibnamefont {Rudenko}}, \bibinfo {author} {\bibfnamefont
  {M.}~\bibnamefont {Seifrid}}, \bibinfo {author} {\bibfnamefont {K.~R.}\
  \bibnamefont {Siefermann}}, \bibinfo {author} {\bibfnamefont {F.~P.}\
  \bibnamefont {Sturm}}, \bibinfo {author} {\bibfnamefont {M.}~\bibnamefont
  {Swiggers}}, \bibinfo {author} {\bibfnamefont {J.}~\bibnamefont {Ullrich}},
  \bibinfo {author} {\bibfnamefont {F.}~\bibnamefont {Weise}}, \bibinfo
  {author} {\bibfnamefont {P.}~\bibnamefont {Zwart}}, \bibinfo {author}
  {\bibfnamefont {C.}~\bibnamefont {Bostedt}}, \bibinfo {author} {\bibfnamefont
  {O.}~\bibnamefont {Gessner}}, \ and\ \bibinfo {author} {\bibfnamefont
  {A.~F.}\ \bibnamefont {Vilesov}},\ }\href {\doibase 10.1126/science.1252395}
  {\bibfield  {journal} {\bibinfo  {journal} {Science}\ }\textbf {\bibinfo
  {volume} {345}},\ \bibinfo {pages} {906} (\bibinfo {year}
  {2014})}\BibitemShut {NoStop}%
\bibitem [{\citenamefont {Stringari}(1999)}]{Stringari1999}%
  \BibitemOpen
  \bibfield  {author} {\bibinfo {author} {\bibfnamefont {S.}~\bibnamefont
  {Stringari}},\ }\href {\doibase 10.1103/PhysRevLett.82.4371} {\bibfield
  {journal} {\bibinfo  {journal} {Phys. Rev. Lett.}\ }\textbf {\bibinfo
  {volume} {82}},\ \bibinfo {pages} {4371} (\bibinfo {year}
  {1999})}\BibitemShut {NoStop}%
\bibitem [{\citenamefont {Butts}\ and\ \citenamefont
  {Rokhsar}(1999)}]{Butts1999}%
  \BibitemOpen
  \bibfield  {author} {\bibinfo {author} {\bibfnamefont {D.~A.}\ \bibnamefont
  {Butts}}\ and\ \bibinfo {author} {\bibfnamefont {D.~S.}\ \bibnamefont
  {Rokhsar}},\ }\href {https://doi.org/10.1038/16865} {\bibfield  {journal}
  {\bibinfo  {journal} {Nature}\ }\textbf {\bibinfo {volume} {397}},\ \bibinfo
  {pages} {327 EP } (\bibinfo {year} {1999})}\BibitemShut {NoStop}%
\bibitem [{\citenamefont {Linn}\ and\ \citenamefont {Fetter}(1999)}]{Linn1999}%
  \BibitemOpen
  \bibfield  {author} {\bibinfo {author} {\bibfnamefont {M.}~\bibnamefont
  {Linn}}\ and\ \bibinfo {author} {\bibfnamefont {A.~L.}\ \bibnamefont
  {Fetter}},\ }\href {\doibase 10.1103/PhysRevA.60.4910} {\bibfield  {journal}
  {\bibinfo  {journal} {Phys. Rev. A}\ }\textbf {\bibinfo {volume} {60}},\
  \bibinfo {pages} {4910} (\bibinfo {year} {1999})}\BibitemShut {NoStop}%
\bibitem [{\citenamefont {Feder}\ \emph {et~al.}(1999)\citenamefont {Feder},
  \citenamefont {Clark},\ and\ \citenamefont {Schneider}}]{Feder1999}%
  \BibitemOpen
  \bibfield  {author} {\bibinfo {author} {\bibfnamefont {D.~L.}\ \bibnamefont
  {Feder}}, \bibinfo {author} {\bibfnamefont {C.~W.}\ \bibnamefont {Clark}}, \
  and\ \bibinfo {author} {\bibfnamefont {B.~I.}\ \bibnamefont {Schneider}},\
  }\href {\doibase 10.1103/PhysRevA.61.011601} {\bibfield  {journal} {\bibinfo
  {journal} {Phys. Rev. A}\ }\textbf {\bibinfo {volume} {61}},\ \bibinfo
  {pages} {011601} (\bibinfo {year} {1999})}\BibitemShut {NoStop}%
\bibitem [{\citenamefont {Kavoulakis}\ \emph {et~al.}(2000)\citenamefont
  {Kavoulakis}, \citenamefont {Mottelson},\ and\ \citenamefont
  {Pethick}}]{Kavoulakis2000}%
  \BibitemOpen
  \bibfield  {author} {\bibinfo {author} {\bibfnamefont {G.~M.}\ \bibnamefont
  {Kavoulakis}}, \bibinfo {author} {\bibfnamefont {B.}~\bibnamefont
  {Mottelson}}, \ and\ \bibinfo {author} {\bibfnamefont {C.~J.}\ \bibnamefont
  {Pethick}},\ }\href {\doibase 10.1103/PhysRevA.62.063605} {\bibfield
  {journal} {\bibinfo  {journal} {Phys. Rev. A}\ }\textbf {\bibinfo {volume}
  {62}},\ \bibinfo {pages} {063605} (\bibinfo {year} {2000})}\BibitemShut
  {NoStop}%
\bibitem [{\citenamefont {Linn}\ \emph {et~al.}(2001)\citenamefont {Linn},
  \citenamefont {Niemeyer},\ and\ \citenamefont {Fetter}}]{Linn2001}%
  \BibitemOpen
  \bibfield  {author} {\bibinfo {author} {\bibfnamefont {M.}~\bibnamefont
  {Linn}}, \bibinfo {author} {\bibfnamefont {M.}~\bibnamefont {Niemeyer}}, \
  and\ \bibinfo {author} {\bibfnamefont {A.~L.}\ \bibnamefont {Fetter}},\
  }\href {\doibase 10.1103/PhysRevA.64.023602} {\bibfield  {journal} {\bibinfo
  {journal} {Phys. Rev. A}\ }\textbf {\bibinfo {volume} {64}},\ \bibinfo
  {pages} {023602} (\bibinfo {year} {2001})}\BibitemShut {NoStop}%
\bibitem [{\citenamefont {Garc\'{\i}a-Ripoll}\ and\ \citenamefont
  {P\'erez-Garc\'{\i}a}(2001)}]{Garcia2001}%
  \BibitemOpen
  \bibfield  {author} {\bibinfo {author} {\bibfnamefont {J.~J.}\ \bibnamefont
  {Garc\'{\i}a-Ripoll}}\ and\ \bibinfo {author} {\bibfnamefont {V.~M.}\
  \bibnamefont {P\'erez-Garc\'{\i}a}},\ }\href {\doibase
  10.1103/PhysRevA.63.041603} {\bibfield  {journal} {\bibinfo  {journal} {Phys.
  Rev. A}\ }\textbf {\bibinfo {volume} {63}},\ \bibinfo {pages} {041603}
  (\bibinfo {year} {2001})}\BibitemShut {NoStop}%
\bibitem [{\citenamefont {Sinha}\ and\ \citenamefont
  {Castin}(2001)}]{Sinha2001}%
  \BibitemOpen
  \bibfield  {author} {\bibinfo {author} {\bibfnamefont {S.}~\bibnamefont
  {Sinha}}\ and\ \bibinfo {author} {\bibfnamefont {Y.}~\bibnamefont {Castin}},\
  }\href {\doibase 10.1103/PhysRevLett.87.190402} {\bibfield  {journal}
  {\bibinfo  {journal} {Phys. Rev. Lett.}\ }\textbf {\bibinfo {volume} {87}},\
  \bibinfo {pages} {190402} (\bibinfo {year} {2001})}\BibitemShut {NoStop}%
\bibitem [{\citenamefont {Kasamatsu}\ \emph {et~al.}(2003)\citenamefont
  {Kasamatsu}, \citenamefont {Tsubota},\ and\ \citenamefont
  {Ueda}}]{Kasamatsu2003}%
  \BibitemOpen
  \bibfield  {author} {\bibinfo {author} {\bibfnamefont {K.}~\bibnamefont
  {Kasamatsu}}, \bibinfo {author} {\bibfnamefont {M.}~\bibnamefont {Tsubota}},
  \ and\ \bibinfo {author} {\bibfnamefont {M.}~\bibnamefont {Ueda}},\ }\href
  {\doibase 10.1103/PhysRevA.67.033610} {\bibfield  {journal} {\bibinfo
  {journal} {Phys. Rev. A}\ }\textbf {\bibinfo {volume} {67}},\ \bibinfo
  {pages} {033610} (\bibinfo {year} {2003})}\BibitemShut {NoStop}%
\bibitem [{\citenamefont {Vorov}\ \emph {et~al.}(2005)\citenamefont {Vorov},
  \citenamefont {Isacker}, \citenamefont {Hussein},\ and\ \citenamefont
  {Bartschat}}]{Vorov2005}%
  \BibitemOpen
  \bibfield  {author} {\bibinfo {author} {\bibfnamefont {O.~K.}\ \bibnamefont
  {Vorov}}, \bibinfo {author} {\bibfnamefont {P.~V.}\ \bibnamefont {Isacker}},
  \bibinfo {author} {\bibfnamefont {M.~S.}\ \bibnamefont {Hussein}}, \ and\
  \bibinfo {author} {\bibfnamefont {K.}~\bibnamefont {Bartschat}},\ }\href
  {\doibase 10.1103/PhysRevLett.95.230406} {\bibfield  {journal} {\bibinfo
  {journal} {Phys. Rev. Lett.}\ }\textbf {\bibinfo {volume} {95}},\ \bibinfo
  {pages} {230406} (\bibinfo {year} {2005})}\BibitemShut {NoStop}%
\bibitem [{\citenamefont {Lieb}\ \emph {et~al.}(2009)\citenamefont {Lieb},
  \citenamefont {Seiringer},\ and\ \citenamefont {Yngvason}}]{Lieb2009}%
  \BibitemOpen
  \bibfield  {author} {\bibinfo {author} {\bibfnamefont {E.~H.}\ \bibnamefont
  {Lieb}}, \bibinfo {author} {\bibfnamefont {R.}~\bibnamefont {Seiringer}}, \
  and\ \bibinfo {author} {\bibfnamefont {J.}~\bibnamefont {Yngvason}},\ }\href
  {\doibase 10.1103/PhysRevA.79.063626} {\bibfield  {journal} {\bibinfo
  {journal} {Phys. Rev. A}\ }\textbf {\bibinfo {volume} {79}},\ \bibinfo
  {pages} {063626} (\bibinfo {year} {2009})}\BibitemShut {NoStop}%
\bibitem [{\citenamefont {Wilkin}\ \emph {et~al.}(1998)\citenamefont {Wilkin},
  \citenamefont {Gunn},\ and\ \citenamefont {Smith}}]{Wilkin1998}%
  \BibitemOpen
  \bibfield  {author} {\bibinfo {author} {\bibfnamefont {N.~K.}\ \bibnamefont
  {Wilkin}}, \bibinfo {author} {\bibfnamefont {J.~M.~F.}\ \bibnamefont {Gunn}},
  \ and\ \bibinfo {author} {\bibfnamefont {R.~A.}\ \bibnamefont {Smith}},\
  }\href {\doibase 10.1103/PhysRevLett.80.2265} {\bibfield  {journal} {\bibinfo
   {journal} {Phys. Rev. Lett.}\ }\textbf {\bibinfo {volume} {80}},\ \bibinfo
  {pages} {2265} (\bibinfo {year} {1998})}\BibitemShut {NoStop}%
\bibitem [{\citenamefont {Mottelson}(1999)}]{Mottelson1999}%
  \BibitemOpen
  \bibfield  {author} {\bibinfo {author} {\bibfnamefont {B.}~\bibnamefont
  {Mottelson}},\ }\href {\doibase 10.1103/PhysRevLett.83.2695} {\bibfield
  {journal} {\bibinfo  {journal} {Phys. Rev. Lett.}\ }\textbf {\bibinfo
  {volume} {83}},\ \bibinfo {pages} {2695} (\bibinfo {year}
  {1999})}\BibitemShut {NoStop}%
\bibitem [{\citenamefont {Bertsch}\ and\ \citenamefont
  {Papenbrock}(1999)}]{Bertsch1999}%
  \BibitemOpen
  \bibfield  {author} {\bibinfo {author} {\bibfnamefont {G.~F.}\ \bibnamefont
  {Bertsch}}\ and\ \bibinfo {author} {\bibfnamefont {T.}~\bibnamefont
  {Papenbrock}},\ }\href {\doibase 10.1103/PhysRevLett.83.5412} {\bibfield
  {journal} {\bibinfo  {journal} {Phys. Rev. Lett.}\ }\textbf {\bibinfo
  {volume} {83}},\ \bibinfo {pages} {5412} (\bibinfo {year}
  {1999})}\BibitemShut {NoStop}%
\bibitem [{\citenamefont {Jackson}\ and\ \citenamefont
  {Kavoulakis}(2000)}]{Jackson2000}%
  \BibitemOpen
  \bibfield  {author} {\bibinfo {author} {\bibfnamefont {A.~D.}\ \bibnamefont
  {Jackson}}\ and\ \bibinfo {author} {\bibfnamefont {G.~M.}\ \bibnamefont
  {Kavoulakis}},\ }\href {\doibase 10.1103/PhysRevLett.85.2854} {\bibfield
  {journal} {\bibinfo  {journal} {Phys. Rev. Lett.}\ }\textbf {\bibinfo
  {volume} {85}},\ \bibinfo {pages} {2854} (\bibinfo {year}
  {2000})}\BibitemShut {NoStop}%
\bibitem [{\citenamefont {Smith}\ and\ \citenamefont
  {Wilkin}(2000)}]{Smith2000}%
  \BibitemOpen
  \bibfield  {author} {\bibinfo {author} {\bibfnamefont {R.~A.}\ \bibnamefont
  {Smith}}\ and\ \bibinfo {author} {\bibfnamefont {N.~K.}\ \bibnamefont
  {Wilkin}},\ }\href {\doibase 10.1103/PhysRevA.62.061602} {\bibfield
  {journal} {\bibinfo  {journal} {Phys. Rev. A}\ }\textbf {\bibinfo {volume}
  {62}},\ \bibinfo {pages} {061602} (\bibinfo {year} {2000})}\BibitemShut
  {NoStop}%
\bibitem [{\citenamefont {Papenbrock}\ and\ \citenamefont
  {Bertsch}(2001)}]{Papenbrock2001}%
  \BibitemOpen
  \bibfield  {author} {\bibinfo {author} {\bibfnamefont {T.}~\bibnamefont
  {Papenbrock}}\ and\ \bibinfo {author} {\bibfnamefont {G.~F.}\ \bibnamefont
  {Bertsch}},\ }\href {\doibase 10.1103/PhysRevA.63.023616} {\bibfield
  {journal} {\bibinfo  {journal} {Phys. Rev. A}\ }\textbf {\bibinfo {volume}
  {63}},\ \bibinfo {pages} {023616} (\bibinfo {year} {2001})}\BibitemShut
  {NoStop}%
\bibitem [{\citenamefont {Huang}(2000)}]{Huang2000}%
  \BibitemOpen
  \bibfield  {author} {\bibinfo {author} {\bibfnamefont {W.-J.}\ \bibnamefont
  {Huang}},\ }\href {\doibase 10.1103/PhysRevA.63.015602} {\bibfield  {journal}
  {\bibinfo  {journal} {Phys. Rev. A}\ }\textbf {\bibinfo {volume} {63}},\
  \bibinfo {pages} {015602} (\bibinfo {year} {2000})}\BibitemShut {NoStop}%
\bibitem [{\citenamefont {Jackson}\ \emph {et~al.}(2001)\citenamefont
  {Jackson}, \citenamefont {Kavoulakis}, \citenamefont {Mottelson},\ and\
  \citenamefont {Reimann}}]{Jackson2001}%
  \BibitemOpen
  \bibfield  {author} {\bibinfo {author} {\bibfnamefont {A.~D.}\ \bibnamefont
  {Jackson}}, \bibinfo {author} {\bibfnamefont {G.~M.}\ \bibnamefont
  {Kavoulakis}}, \bibinfo {author} {\bibfnamefont {B.}~\bibnamefont
  {Mottelson}}, \ and\ \bibinfo {author} {\bibfnamefont {S.~M.}\ \bibnamefont
  {Reimann}},\ }\href {\doibase 10.1103/PhysRevLett.86.945} {\bibfield
  {journal} {\bibinfo  {journal} {Phys. Rev. Lett.}\ }\textbf {\bibinfo
  {volume} {86}},\ \bibinfo {pages} {945} (\bibinfo {year} {2001})}\BibitemShut
  {NoStop}%
\bibitem [{\citenamefont {Liu}\ \emph {et~al.}(2001)\citenamefont {Liu},
  \citenamefont {Hu}, \citenamefont {Chang}, \citenamefont {Zhang},
  \citenamefont {Li},\ and\ \citenamefont {Wang}}]{Liu2001}%
  \BibitemOpen
  \bibfield  {author} {\bibinfo {author} {\bibfnamefont {X.-J.}\ \bibnamefont
  {Liu}}, \bibinfo {author} {\bibfnamefont {H.}~\bibnamefont {Hu}}, \bibinfo
  {author} {\bibfnamefont {L.}~\bibnamefont {Chang}}, \bibinfo {author}
  {\bibfnamefont {W.}~\bibnamefont {Zhang}}, \bibinfo {author} {\bibfnamefont
  {S.-Q.}\ \bibnamefont {Li}}, \ and\ \bibinfo {author} {\bibfnamefont {Y.-Z.}\
  \bibnamefont {Wang}},\ }\href {\doibase 10.1103/PhysRevLett.87.030404}
  {\bibfield  {journal} {\bibinfo  {journal} {Phys. Rev. Lett.}\ }\textbf
  {\bibinfo {volume} {87}},\ \bibinfo {pages} {030404} (\bibinfo {year}
  {2001})}\BibitemShut {NoStop}%
\bibitem [{\citenamefont {Tsubota}\ \emph {et~al.}(2002)\citenamefont
  {Tsubota}, \citenamefont {Kasamatsu},\ and\ \citenamefont
  {Ueda}}]{Tsubota2002}%
  \BibitemOpen
  \bibfield  {author} {\bibinfo {author} {\bibfnamefont {M.}~\bibnamefont
  {Tsubota}}, \bibinfo {author} {\bibfnamefont {K.}~\bibnamefont {Kasamatsu}},
  \ and\ \bibinfo {author} {\bibfnamefont {M.}~\bibnamefont {Ueda}},\ }\href
  {\doibase 10.1103/PhysRevA.65.023603} {\bibfield  {journal} {\bibinfo
  {journal} {Phys. Rev. A}\ }\textbf {\bibinfo {volume} {65}},\ \bibinfo
  {pages} {023603} (\bibinfo {year} {2002})}\BibitemShut {NoStop}%
\bibitem [{\citenamefont {Manninen}\ \emph {et~al.}(2005)\citenamefont
  {Manninen}, \citenamefont {Reimann}, \citenamefont {Koskinen}, \citenamefont
  {Yu},\ and\ \citenamefont {Toreblad}}]{Manninen2005}%
  \BibitemOpen
  \bibfield  {author} {\bibinfo {author} {\bibfnamefont {M.}~\bibnamefont
  {Manninen}}, \bibinfo {author} {\bibfnamefont {S.~M.}\ \bibnamefont
  {Reimann}}, \bibinfo {author} {\bibfnamefont {M.}~\bibnamefont {Koskinen}},
  \bibinfo {author} {\bibfnamefont {Y.}~\bibnamefont {Yu}}, \ and\ \bibinfo
  {author} {\bibfnamefont {M.}~\bibnamefont {Toreblad}},\ }\href {\doibase
  10.1103/PhysRevLett.94.106405} {\bibfield  {journal} {\bibinfo  {journal}
  {Phys. Rev. Lett.}\ }\textbf {\bibinfo {volume} {94}},\ \bibinfo {pages}
  {106405} (\bibinfo {year} {2005})}\BibitemShut {NoStop}%
\bibitem [{\citenamefont {Reimann}\ \emph {et~al.}(2006)\citenamefont
  {Reimann}, \citenamefont {Koskinen}, \citenamefont {Yu},\ and\ \citenamefont
  {Manninen}}]{Reimann2006}%
  \BibitemOpen
  \bibfield  {author} {\bibinfo {author} {\bibfnamefont {S.~M.}\ \bibnamefont
  {Reimann}}, \bibinfo {author} {\bibfnamefont {M.}~\bibnamefont {Koskinen}},
  \bibinfo {author} {\bibfnamefont {Y.}~\bibnamefont {Yu}}, \ and\ \bibinfo
  {author} {\bibfnamefont {M.}~\bibnamefont {Manninen}},\ }\href {\doibase
  10.1103/PhysRevA.74.043603} {\bibfield  {journal} {\bibinfo  {journal} {Phys.
  Rev. A}\ }\textbf {\bibinfo {volume} {74}},\ \bibinfo {pages} {043603}
  (\bibinfo {year} {2006})}\BibitemShut {NoStop}%
\bibitem [{\citenamefont {Barber\'an}\ \emph {et~al.}(2006)\citenamefont
  {Barber\'an}, \citenamefont {Lewenstein}, \citenamefont {Osterloh},\ and\
  \citenamefont {Dagnino}}]{Barberan2006}%
  \BibitemOpen
  \bibfield  {author} {\bibinfo {author} {\bibfnamefont {N.}~\bibnamefont
  {Barber\'an}}, \bibinfo {author} {\bibfnamefont {M.}~\bibnamefont
  {Lewenstein}}, \bibinfo {author} {\bibfnamefont {K.}~\bibnamefont
  {Osterloh}}, \ and\ \bibinfo {author} {\bibfnamefont {D.}~\bibnamefont
  {Dagnino}},\ }\href {\doibase 10.1103/PhysRevA.73.063623} {\bibfield
  {journal} {\bibinfo  {journal} {Phys. Rev. A}\ }\textbf {\bibinfo {volume}
  {73}},\ \bibinfo {pages} {063623} (\bibinfo {year} {2006})}\BibitemShut
  {NoStop}%
\bibitem [{\citenamefont {Dagnino}\ \emph {et~al.}(2007)\citenamefont
  {Dagnino}, \citenamefont {Barber\'an}, \citenamefont {Osterloh},
  \citenamefont {Riera},\ and\ \citenamefont {Lewenstein}}]{Dagnino2007}%
  \BibitemOpen
  \bibfield  {author} {\bibinfo {author} {\bibfnamefont {D.}~\bibnamefont
  {Dagnino}}, \bibinfo {author} {\bibfnamefont {N.}~\bibnamefont {Barber\'an}},
  \bibinfo {author} {\bibfnamefont {K.}~\bibnamefont {Osterloh}}, \bibinfo
  {author} {\bibfnamefont {A.}~\bibnamefont {Riera}}, \ and\ \bibinfo {author}
  {\bibfnamefont {M.}~\bibnamefont {Lewenstein}},\ }\href {\doibase
  10.1103/PhysRevA.76.013625} {\bibfield  {journal} {\bibinfo  {journal} {Phys.
  Rev. A}\ }\textbf {\bibinfo {volume} {76}},\ \bibinfo {pages} {013625}
  (\bibinfo {year} {2007})}\BibitemShut {NoStop}%
\bibitem [{\citenamefont {Cooper}(2008)}]{Cooper2008}%
  \BibitemOpen
  \bibfield  {author} {\bibinfo {author} {\bibfnamefont {N.}~\bibnamefont
  {Cooper}},\ }\href {\doibase 10.1080/00018730802564122} {\bibfield  {journal}
  {\bibinfo  {journal} {Advances in Physics}\ }\textbf {\bibinfo {volume}
  {57}},\ \bibinfo {pages} {539} (\bibinfo {year} {2008})}\BibitemShut
  {NoStop}%
\bibitem [{\citenamefont {Parke}\ \emph {et~al.}(2008)\citenamefont {Parke},
  \citenamefont {Wilkin}, \citenamefont {Gunn},\ and\ \citenamefont
  {Bourne}}]{Parke2008}%
  \BibitemOpen
  \bibfield  {author} {\bibinfo {author} {\bibfnamefont {M.~I.}\ \bibnamefont
  {Parke}}, \bibinfo {author} {\bibfnamefont {N.~K.}\ \bibnamefont {Wilkin}},
  \bibinfo {author} {\bibfnamefont {J.~M.~F.}\ \bibnamefont {Gunn}}, \ and\
  \bibinfo {author} {\bibfnamefont {A.}~\bibnamefont {Bourne}},\ }\href
  {\doibase 10.1103/PhysRevLett.101.110401} {\bibfield  {journal} {\bibinfo
  {journal} {Phys. Rev. Lett.}\ }\textbf {\bibinfo {volume} {101}},\ \bibinfo
  {pages} {110401} (\bibinfo {year} {2008})}\BibitemShut {NoStop}%
\bibitem [{\citenamefont {Romanovsky}\ \emph {et~al.}(2008)\citenamefont
  {Romanovsky}, \citenamefont {Yannouleas},\ and\ \citenamefont
  {Landman}}]{Romanovsky2008}%
  \BibitemOpen
  \bibfield  {author} {\bibinfo {author} {\bibfnamefont {I.}~\bibnamefont
  {Romanovsky}}, \bibinfo {author} {\bibfnamefont {C.}~\bibnamefont
  {Yannouleas}}, \ and\ \bibinfo {author} {\bibfnamefont {U.}~\bibnamefont
  {Landman}},\ }\href {\doibase 10.1103/PhysRevA.78.011606} {\bibfield
  {journal} {\bibinfo  {journal} {Phys. Rev. A}\ }\textbf {\bibinfo {volume}
  {78}},\ \bibinfo {pages} {011606} (\bibinfo {year} {2008})}\BibitemShut
  {NoStop}%
\bibitem [{\citenamefont {Liu}\ \emph {et~al.}(2009)\citenamefont {Liu},
  \citenamefont {Guo}, \citenamefont {Chen},\ and\ \citenamefont
  {Fan}}]{Liu2009}%
  \BibitemOpen
  \bibfield  {author} {\bibinfo {author} {\bibfnamefont {Z.}~\bibnamefont
  {Liu}}, \bibinfo {author} {\bibfnamefont {H.}~\bibnamefont {Guo}}, \bibinfo
  {author} {\bibfnamefont {S.}~\bibnamefont {Chen}}, \ and\ \bibinfo {author}
  {\bibfnamefont {H.}~\bibnamefont {Fan}},\ }\href {\doibase
  10.1103/PhysRevA.80.063606} {\bibfield  {journal} {\bibinfo  {journal} {Phys.
  Rev. A}\ }\textbf {\bibinfo {volume} {80}},\ \bibinfo {pages} {063606}
  (\bibinfo {year} {2009})}\BibitemShut {NoStop}%
\bibitem [{\citenamefont {Dagnino}\ \emph
  {et~al.}(2009{\natexlab{b}})\citenamefont {Dagnino}, \citenamefont
  {Barber\'an},\ and\ \citenamefont {Lewenstein}}]{Dagnino2009b}%
  \BibitemOpen
  \bibfield  {author} {\bibinfo {author} {\bibfnamefont {D.}~\bibnamefont
  {Dagnino}}, \bibinfo {author} {\bibfnamefont {N.}~\bibnamefont {Barber\'an}},
  \ and\ \bibinfo {author} {\bibfnamefont {M.}~\bibnamefont {Lewenstein}},\
  }\href {\doibase 10.1103/PhysRevA.80.053611} {\bibfield  {journal} {\bibinfo
  {journal} {Phys. Rev. A}\ }\textbf {\bibinfo {volume} {80}},\ \bibinfo
  {pages} {053611} (\bibinfo {year} {2009}{\natexlab{b}})}\BibitemShut
  {NoStop}%
\bibitem [{\citenamefont {Papenbrock}\ \emph {et~al.}(2012)\citenamefont
  {Papenbrock}, \citenamefont {Reimann},\ and\ \citenamefont
  {Kavoulakis}}]{Papenbrock2012}%
  \BibitemOpen
  \bibfield  {author} {\bibinfo {author} {\bibfnamefont {T.}~\bibnamefont
  {Papenbrock}}, \bibinfo {author} {\bibfnamefont {S.~M.}\ \bibnamefont
  {Reimann}}, \ and\ \bibinfo {author} {\bibfnamefont {G.~M.}\ \bibnamefont
  {Kavoulakis}},\ }\href {\doibase 10.1103/PhysRevLett.108.075304} {\bibfield
  {journal} {\bibinfo  {journal} {Phys. Rev. Lett.}\ }\textbf {\bibinfo
  {volume} {108}},\ \bibinfo {pages} {075304} (\bibinfo {year}
  {2012})}\BibitemShut {NoStop}%
\bibitem [{\citenamefont {Cremon}\ \emph {et~al.}(2013)\citenamefont {Cremon},
  \citenamefont {Kavoulakis}, \citenamefont {Mottelson},\ and\ \citenamefont
  {Reimann}}]{Cremon2013}%
  \BibitemOpen
  \bibfield  {author} {\bibinfo {author} {\bibfnamefont {J.~C.}\ \bibnamefont
  {Cremon}}, \bibinfo {author} {\bibfnamefont {G.~M.}\ \bibnamefont
  {Kavoulakis}}, \bibinfo {author} {\bibfnamefont {B.~R.}\ \bibnamefont
  {Mottelson}}, \ and\ \bibinfo {author} {\bibfnamefont {S.~M.}\ \bibnamefont
  {Reimann}},\ }\href {\doibase 10.1103/PhysRevA.87.053615} {\bibfield
  {journal} {\bibinfo  {journal} {Phys. Rev. A}\ }\textbf {\bibinfo {volume}
  {87}},\ \bibinfo {pages} {053615} (\bibinfo {year} {2013})}\BibitemShut
  {NoStop}%
\bibitem [{\citenamefont {Cremon}\ \emph {et~al.}(2015)\citenamefont {Cremon},
  \citenamefont {Jackson}, \citenamefont {Karabulut}, \citenamefont
  {Kavoulakis}, \citenamefont {Mottelson},\ and\ \citenamefont
  {Reimann}}]{Cremon2015}%
  \BibitemOpen
  \bibfield  {author} {\bibinfo {author} {\bibfnamefont {J.~C.}\ \bibnamefont
  {Cremon}}, \bibinfo {author} {\bibfnamefont {A.~D.}\ \bibnamefont {Jackson}},
  \bibinfo {author} {\bibfnamefont {E.~O.}\ \bibnamefont {Karabulut}}, \bibinfo
  {author} {\bibfnamefont {G.~M.}\ \bibnamefont {Kavoulakis}}, \bibinfo
  {author} {\bibfnamefont {B.~R.}\ \bibnamefont {Mottelson}}, \ and\ \bibinfo
  {author} {\bibfnamefont {S.~M.}\ \bibnamefont {Reimann}},\ }\href {\doibase
  10.1103/PhysRevA.91.033623} {\bibfield  {journal} {\bibinfo  {journal} {Phys.
  Rev. A}\ }\textbf {\bibinfo {volume} {91}},\ \bibinfo {pages} {033623}
  (\bibinfo {year} {2015})}\BibitemShut {NoStop}%
\bibitem [{\citenamefont {Weiner}\ \emph {et~al.}(2017)\citenamefont {Weiner},
  \citenamefont {Tsatsos}, \citenamefont {Cederbaum},\ and\ \citenamefont
  {Lode}}]{Weiner2017}%
  \BibitemOpen
  \bibfield  {author} {\bibinfo {author} {\bibfnamefont {S.~E.}\ \bibnamefont
  {Weiner}}, \bibinfo {author} {\bibfnamefont {M.~C.}\ \bibnamefont {Tsatsos}},
  \bibinfo {author} {\bibfnamefont {L.~S.}\ \bibnamefont {Cederbaum}}, \ and\
  \bibinfo {author} {\bibfnamefont {A.~U.}\ \bibnamefont {Lode}},\ }\href@noop
  {} {\bibfield  {journal} {\bibinfo  {journal} {Scientific reports}\ }\textbf
  {\bibinfo {volume} {7}},\ \bibinfo {pages} {40122} (\bibinfo {year}
  {2017})}\BibitemShut {NoStop}%
\bibitem [{\citenamefont {Beinke}\ \emph {et~al.}(2018)\citenamefont {Beinke},
  \citenamefont {Cederbaum},\ and\ \citenamefont {Alon}}]{Beinke2018}%
  \BibitemOpen
  \bibfield  {author} {\bibinfo {author} {\bibfnamefont {R.}~\bibnamefont
  {Beinke}}, \bibinfo {author} {\bibfnamefont {L.~S.}\ \bibnamefont
  {Cederbaum}}, \ and\ \bibinfo {author} {\bibfnamefont {O.~E.}\ \bibnamefont
  {Alon}},\ }\href {\doibase 10.1103/PhysRevA.98.053634} {\bibfield  {journal}
  {\bibinfo  {journal} {Phys. Rev. A}\ }\textbf {\bibinfo {volume} {98}},\
  \bibinfo {pages} {053634} (\bibinfo {year} {2018})}\BibitemShut {NoStop}%
\bibitem [{\citenamefont {Cooper}\ and\ \citenamefont
  {Wilkin}(1999)}]{Cooper1999}%
  \BibitemOpen
  \bibfield  {author} {\bibinfo {author} {\bibfnamefont {N.~R.}\ \bibnamefont
  {Cooper}}\ and\ \bibinfo {author} {\bibfnamefont {N.~K.}\ \bibnamefont
  {Wilkin}},\ }\href {\doibase 10.1103/PhysRevB.60.R16279} {\bibfield
  {journal} {\bibinfo  {journal} {Phys. Rev. B}\ }\textbf {\bibinfo {volume}
  {60}},\ \bibinfo {pages} {R16279} (\bibinfo {year} {1999})}\BibitemShut
  {NoStop}%
\bibitem [{\citenamefont {Wilkin}\ and\ \citenamefont
  {Gunn}(2000)}]{Wilkin2000}%
  \BibitemOpen
  \bibfield  {author} {\bibinfo {author} {\bibfnamefont {N.~K.}\ \bibnamefont
  {Wilkin}}\ and\ \bibinfo {author} {\bibfnamefont {J.~M.~F.}\ \bibnamefont
  {Gunn}},\ }\href {\doibase 10.1103/PhysRevLett.84.6} {\bibfield  {journal}
  {\bibinfo  {journal} {Phys. Rev. Lett.}\ }\textbf {\bibinfo {volume} {84}},\
  \bibinfo {pages} {6} (\bibinfo {year} {2000})}\BibitemShut {NoStop}%
\bibitem [{\citenamefont {Viefers}\ \emph {et~al.}(2000)\citenamefont
  {Viefers}, \citenamefont {Hansson},\ and\ \citenamefont
  {Reimann}}]{Viefers2000}%
  \BibitemOpen
  \bibfield  {author} {\bibinfo {author} {\bibfnamefont {S.}~\bibnamefont
  {Viefers}}, \bibinfo {author} {\bibfnamefont {T.~H.}\ \bibnamefont
  {Hansson}}, \ and\ \bibinfo {author} {\bibfnamefont {S.~M.}\ \bibnamefont
  {Reimann}},\ }\href {\doibase 10.1103/PhysRevA.62.053604} {\bibfield
  {journal} {\bibinfo  {journal} {Phys. Rev. A}\ }\textbf {\bibinfo {volume}
  {62}},\ \bibinfo {pages} {053604} (\bibinfo {year} {2000})}\BibitemShut
  {NoStop}%
\bibitem [{\citenamefont {Viefers}(2008)}]{Viefers2008}%
  \BibitemOpen
  \bibfield  {author} {\bibinfo {author} {\bibfnamefont {S.}~\bibnamefont
  {Viefers}},\ }\href {http://stacks.iop.org/0953-8984/20/i=12/a=123202}
  {\bibfield  {journal} {\bibinfo  {journal} {Journal of Physics: Condensed
  Matter}\ }\textbf {\bibinfo {volume} {20}},\ \bibinfo {pages} {123202}
  (\bibinfo {year} {2008})}\BibitemShut {NoStop}%
\bibitem [{\citenamefont {Bloch}\ \emph {et~al.}(2008)\citenamefont {Bloch},
  \citenamefont {Dalibard},\ and\ \citenamefont {Zwerger}}]{Bloch2008}%
  \BibitemOpen
  \bibfield  {author} {\bibinfo {author} {\bibfnamefont {I.}~\bibnamefont
  {Bloch}}, \bibinfo {author} {\bibfnamefont {J.}~\bibnamefont {Dalibard}}, \
  and\ \bibinfo {author} {\bibfnamefont {W.}~\bibnamefont {Zwerger}},\ }\href
  {\doibase 10.1103/RevModPhys.80.885} {\bibfield  {journal} {\bibinfo
  {journal} {Rev. Mod. Phys.}\ }\textbf {\bibinfo {volume} {80}},\ \bibinfo
  {pages} {885} (\bibinfo {year} {2008})}\BibitemShut {NoStop}%
\bibitem [{\citenamefont {Fetter}(2009)}]{Fetter2009}%
  \BibitemOpen
  \bibfield  {author} {\bibinfo {author} {\bibfnamefont {A.~L.}\ \bibnamefont
  {Fetter}},\ }\href {\doibase 10.1103/RevModPhys.81.647} {\bibfield  {journal}
  {\bibinfo  {journal} {Rev. Mod. Phys.}\ }\textbf {\bibinfo {volume} {81}},\
  \bibinfo {pages} {647} (\bibinfo {year} {2009})}\BibitemShut {NoStop}%
\bibitem [{\citenamefont {Saarikoski}\ \emph {et~al.}(2010)\citenamefont
  {Saarikoski}, \citenamefont {Reimann}, \citenamefont {Harju},\ and\
  \citenamefont {Manninen}}]{Saarikoski2010}%
  \BibitemOpen
  \bibfield  {author} {\bibinfo {author} {\bibfnamefont {H.}~\bibnamefont
  {Saarikoski}}, \bibinfo {author} {\bibfnamefont {S.~M.}\ \bibnamefont
  {Reimann}}, \bibinfo {author} {\bibfnamefont {A.}~\bibnamefont {Harju}}, \
  and\ \bibinfo {author} {\bibfnamefont {M.}~\bibnamefont {Manninen}},\ }\href
  {\doibase 10.1103/RevModPhys.82.2785} {\bibfield  {journal} {\bibinfo
  {journal} {Rev. Mod. Phys.}\ }\textbf {\bibinfo {volume} {82}},\ \bibinfo
  {pages} {2785} (\bibinfo {year} {2010})}\BibitemShut {NoStop}%
\bibitem [{\citenamefont {Ueda}\ and\ \citenamefont
  {Nakajima}(2006)}]{Ueda2006}%
  \BibitemOpen
  \bibfield  {author} {\bibinfo {author} {\bibfnamefont {M.}~\bibnamefont
  {Ueda}}\ and\ \bibinfo {author} {\bibfnamefont {T.}~\bibnamefont
  {Nakajima}},\ }\href {\doibase 10.1103/PhysRevA.73.043603} {\bibfield
  {journal} {\bibinfo  {journal} {Phys. Rev. A}\ }\textbf {\bibinfo {volume}
  {73}},\ \bibinfo {pages} {043603} (\bibinfo {year} {2006})}\BibitemShut
  {NoStop}%
\bibitem [{\citenamefont {Penrose}(1951)}]{Penrose1951}%
  \BibitemOpen
  \bibfield  {author} {\bibinfo {author} {\bibfnamefont {O.}~\bibnamefont
  {Penrose}},\ }\href@noop {} {\bibfield  {journal} {\bibinfo  {journal} {Phil.
  Mag.}\ }\textbf {\bibinfo {volume} {42}},\ \bibinfo {pages} {1373} (\bibinfo
  {year} {1951})}\BibitemShut {NoStop}%
\bibitem [{\citenamefont {Penrose}\ and\ \citenamefont
  {Onsager}(1956)}]{Penrose1956}%
  \BibitemOpen
  \bibfield  {author} {\bibinfo {author} {\bibfnamefont {O.}~\bibnamefont
  {Penrose}}\ and\ \bibinfo {author} {\bibfnamefont {L.}~\bibnamefont
  {Onsager}},\ }\href {\doibase 10.1103/PhysRev.104.576} {\bibfield  {journal}
  {\bibinfo  {journal} {Phys. Rev.}\ }\textbf {\bibinfo {volume} {104}},\
  \bibinfo {pages} {576} (\bibinfo {year} {1956})}\BibitemShut {NoStop}%
\bibitem [{\citenamefont {Yang}(1962)}]{Yang1962}%
  \BibitemOpen
  \bibfield  {author} {\bibinfo {author} {\bibfnamefont {C.~N.}\ \bibnamefont
  {Yang}},\ }\href {\doibase 10.1103/RevModPhys.34.694} {\bibfield  {journal}
  {\bibinfo  {journal} {Rev. Mod. Phys.}\ }\textbf {\bibinfo {volume} {34}},\
  \bibinfo {pages} {694} (\bibinfo {year} {1962})}\BibitemShut {NoStop}%
\bibitem [{\citenamefont {Rico-Gutierrez}\ \emph {et~al.}(2013)\citenamefont
  {Rico-Gutierrez}, \citenamefont {Spiller},\ and\ \citenamefont
  {Dunningham}}]{rico2013}%
  \BibitemOpen
  \bibfield  {author} {\bibinfo {author} {\bibfnamefont {L.~M.}\ \bibnamefont
  {Rico-Gutierrez}}, \bibinfo {author} {\bibfnamefont {T.~P.}\ \bibnamefont
  {Spiller}}, \ and\ \bibinfo {author} {\bibfnamefont {J.~A.}\ \bibnamefont
  {Dunningham}},\ }\href {\doibase 10.1088/1367-2630/15/6/063010} {\bibfield
  {journal} {\bibinfo  {journal} {New Journal of Physics}\ }\textbf {\bibinfo
  {volume} {15}},\ \bibinfo {pages} {063010} (\bibinfo {year}
  {2013})}\BibitemShut {NoStop}%
\bibitem [{\citenamefont {Fr\"owis}\ \emph {et~al.}(2018)\citenamefont
  {Fr\"owis}, \citenamefont {Sekatski}, \citenamefont {D\"ur}, \citenamefont
  {Gisin},\ and\ \citenamefont {Sangouard}}]{Frowis2018}%
  \BibitemOpen
  \bibfield  {author} {\bibinfo {author} {\bibfnamefont {F.}~\bibnamefont
  {Fr\"owis}}, \bibinfo {author} {\bibfnamefont {P.}~\bibnamefont {Sekatski}},
  \bibinfo {author} {\bibfnamefont {W.}~\bibnamefont {D\"ur}}, \bibinfo
  {author} {\bibfnamefont {N.}~\bibnamefont {Gisin}}, \ and\ \bibinfo {author}
  {\bibfnamefont {N.}~\bibnamefont {Sangouard}},\ }\href {\doibase
  10.1103/RevModPhys.90.025004} {\bibfield  {journal} {\bibinfo  {journal}
  {Rev. Mod. Phys.}\ }\textbf {\bibinfo {volume} {90}},\ \bibinfo {pages}
  {025004} (\bibinfo {year} {2018})}\BibitemShut {NoStop}%
\bibitem [{\citenamefont {Ho}(2001)}]{Ho2001}%
  \BibitemOpen
  \bibfield  {author} {\bibinfo {author} {\bibfnamefont {T.-L.}\ \bibnamefont
  {Ho}},\ }\href {\doibase 10.1103/PhysRevLett.87.060403} {\bibfield  {journal}
  {\bibinfo  {journal} {Phys. Rev. Lett.}\ }\textbf {\bibinfo {volume} {87}},\
  \bibinfo {pages} {060403} (\bibinfo {year} {2001})}\BibitemShut {NoStop}%
\bibitem [{\citenamefont {Morris}\ and\ \citenamefont
  {Feder}(2006)}]{Morris2006}%
  \BibitemOpen
  \bibfield  {author} {\bibinfo {author} {\bibfnamefont {A.~G.}\ \bibnamefont
  {Morris}}\ and\ \bibinfo {author} {\bibfnamefont {D.~L.}\ \bibnamefont
  {Feder}},\ }\href {\doibase 10.1103/PhysRevA.74.033605} {\bibfield  {journal}
  {\bibinfo  {journal} {Phys. Rev. A}\ }\textbf {\bibinfo {volume} {74}},\
  \bibinfo {pages} {033605} (\bibinfo {year} {2006})}\BibitemShut {NoStop}%
\bibitem [{\citenamefont {Rontani}\ \emph {et~al.}(2017)\citenamefont
  {Rontani}, \citenamefont {Eriksson}, \citenamefont {{\AA}berg},\ and\
  \citenamefont {Reimann}}]{Rontani2017}%
  \BibitemOpen
  \bibfield  {author} {\bibinfo {author} {\bibfnamefont {M.}~\bibnamefont
  {Rontani}}, \bibinfo {author} {\bibfnamefont {G.}~\bibnamefont {Eriksson}},
  \bibinfo {author} {\bibfnamefont {S.}~\bibnamefont {{\AA}berg}}, \ and\
  \bibinfo {author} {\bibfnamefont {S.}~\bibnamefont {Reimann}},\ }\href
  {\doibase 10.1088/1361-6455/aa606a} {\bibfield  {journal} {\bibinfo
  {journal} {Journal of Optics B: Quantum and Semiclassical Optics}\ }\textbf
  {\bibinfo {volume} {50}} (\bibinfo {year} {2017}),\
  10.1088/1361-6455/aa606a}\BibitemShut {NoStop}%
\bibitem [{\citenamefont {Lipkin}\ \emph {et~al.}(1965)\citenamefont {Lipkin},
  \citenamefont {Meshkov},\ and\ \citenamefont {Glick}}]{Lipkin1965}%
  \BibitemOpen
  \bibfield  {author} {\bibinfo {author} {\bibfnamefont {H.}~\bibnamefont
  {Lipkin}}, \bibinfo {author} {\bibfnamefont {N.}~\bibnamefont {Meshkov}}, \
  and\ \bibinfo {author} {\bibfnamefont {A.}~\bibnamefont {Glick}},\ }\href
  {\doibase https://doi.org/10.1016/0029-5582(65)90862-X} {\bibfield  {journal}
  {\bibinfo  {journal} {Nuclear Physics}\ }\textbf {\bibinfo {volume} {62}},\
  \bibinfo {pages} {188 } (\bibinfo {year} {1965})}\BibitemShut {NoStop}%
\bibitem [{\citenamefont {Elgar\o{}y}\ and\ \citenamefont
  {Pethick}(1999)}]{Elgaroy1999}%
  \BibitemOpen
  \bibfield  {author} {\bibinfo {author} {\bibfnamefont {O.}~\bibnamefont
  {Elgar\o{}y}}\ and\ \bibinfo {author} {\bibfnamefont {C.~J.}\ \bibnamefont
  {Pethick}},\ }\href {\doibase 10.1103/PhysRevA.59.1711} {\bibfield  {journal}
  {\bibinfo  {journal} {Phys. Rev. A}\ }\textbf {\bibinfo {volume} {59}},\
  \bibinfo {pages} {1711} (\bibinfo {year} {1999})}\BibitemShut {NoStop}%
\bibitem [{\citenamefont {Juli\'a-D\'{\i}az}\ \emph {et~al.}(2013)\citenamefont
  {Juli\'a-D\'{\i}az}, \citenamefont {Gottlieb}, \citenamefont {Martorell},\
  and\ \citenamefont {Polls}}]{JuliaDiaz2013}%
  \BibitemOpen
  \bibfield  {author} {\bibinfo {author} {\bibfnamefont {B.}~\bibnamefont
  {Juli\'a-D\'{\i}az}}, \bibinfo {author} {\bibfnamefont {A.~D.}\ \bibnamefont
  {Gottlieb}}, \bibinfo {author} {\bibfnamefont {J.}~\bibnamefont {Martorell}},
  \ and\ \bibinfo {author} {\bibfnamefont {A.}~\bibnamefont {Polls}},\ }\href
  {\doibase 10.1103/PhysRevA.88.033601} {\bibfield  {journal} {\bibinfo
  {journal} {Phys. Rev. A}\ }\textbf {\bibinfo {volume} {88}},\ \bibinfo
  {pages} {033601} (\bibinfo {year} {2013})}\BibitemShut {NoStop}%
\bibitem [{\citenamefont {{Nozi\`eres, P.}}\ and\ \citenamefont {{Saint James,
  D.}}(1982)}]{Nozieres1982}%
  \BibitemOpen
  \bibfield  {author} {\bibinfo {author} {\bibnamefont {{Nozi\`eres, P.}}}\
  and\ \bibinfo {author} {\bibnamefont {{Saint James, D.}}},\ }\href {\doibase
  10.1051/jphys:019820043070113300} {\bibfield  {journal} {\bibinfo  {journal}
  {J. Phys. France}\ }\textbf {\bibinfo {volume} {43}},\ \bibinfo {pages}
  {1133} (\bibinfo {year} {1982})}\BibitemShut {NoStop}%
\bibitem [{\citenamefont {Anderson}\ and\ \citenamefont
  {Bj{\"o}rck}(1973)}]{Anderson1973}%
  \BibitemOpen
  \bibfield  {author} {\bibinfo {author} {\bibfnamefont {N.}~\bibnamefont
  {Anderson}}\ and\ \bibinfo {author} {\bibfnamefont {{\AA}.}~\bibnamefont
  {Bj{\"o}rck}},\ }\href {\doibase 10.1007/BF01951936} {\bibfield  {journal}
  {\bibinfo  {journal} {BIT Numerical Mathematics}\ }\textbf {\bibinfo {volume}
  {13}},\ \bibinfo {pages} {253} (\bibinfo {year} {1973})}\BibitemShut
  {NoStop}%
\end{thebibliography}
\end{document}